\documentclass[12pt, preprint, authoryear]{elsarticle}	
\RequirePackage[colorlinks,citecolor=blue,urlcolor=blue]{hyperref}
\usepackage{amsmath,bm}
\usepackage{amsthm}
\usepackage{amsopn}
\usepackage{graphicx}
\usepackage{amssymb}
\usepackage{amsfonts}
\usepackage[usenames,dvipsnames]{color}
\usepackage[total={150mm,240mm}]{geometry}

\journal{Environmetrics}
 \graphicspath{{plot}{plot/PDF/}{plot/}}
\begin{document}
	\begin{frontmatter}
		\title{Regularized Spatial Maximum Covariance Analysis}
	
	\author[rvt]{Wen-Ting Wang}
	\ead{egpivo@gmail.com}
	
	\author[focal]{Hsin-Cheng Huang\corref{cor1}}
	\cortext[cor1]{Corresponding author}
	\ead{hchuang@stat.sinica.edu.tw}
	
	\address[rvt]{Institute of Statistics, National Chiao Tung University}
	\address[focal]{Institute of Statistical Science, Academia Sinica}
	
		\begin{abstract}
			In climate and atmospheric research, many phenomena involve more than one meteorological spatial processes covarying in space. To understand how one process is affected by another, maximum covariance analysis (MCA) is commonly applied. However, the patterns obtained from MCA may sometimes be difficult to interpret. In this paper,  we propose a regularization approach to promote spatial features in dominant coupled patterns by introducing smoothness and  sparseness penalties while accounting for their orthogonalities. We develop an efficient algorithm to solve the resulting optimization problem by using the alternating direction method of multipliers. The effectiveness of the proposed method is illustrated by several numerical examples, including an application to study how precipitations in east Africa are affected by sea surface temperatures in the Indian Ocean.
		\end{abstract}	
	\end{frontmatter}
\section{Introduction}
Many climate and atmospheric phenomena involve more than one meteorological spatial processes covarying in space. It is of interest to find dominant coupled patterns among these processes. For example, variations of sea surface temperatures (SSTs) in the Indian Ocean may affect precipitations in nearby countries in Africa, particularly over sensitive agricultural regions, and hence threaten the economies and livelihoods of these countries. Consequently, many studies have been conducted on the relationship between the SST and precipitation by analyzing their coupled patterns \citep[e.g.][]{reason2002sensitivity,morioka2012subtropical,omondi2013influence}. A commonly used method is maximum covariance analysis (MCA), which seeks important spatial patterns that explain the maximum amount of covariance between the two processes by using the singular value decomposition (SVD) of the cross-covariance matrix \citep{tucker1958inter}. 

However, the leading coupled patterns obtained by MCA may sometimes be too noisy to be physically interpretable when the signal-to-noise ratio is low. Many approaches have been proposed to improve MCA. For example, \citet{salim2005modelling} and \citet{salim2007model} proposed penalized likelihood approaches using roughness penalties to promote smoothness of the leading coupled patterns in space. However, these methods tend to capture global features but not localized ones. On the other hand, \citet{witten2009penalized} considered canonical correlation analysis with an $L_1$ constraint, and \citet{lee2011sparse} proposed a penalized likelihood method with the SCAD penalty \citep{fan2001variable} to facilitate sparse patterns. However, these methods cannot be applied to continuous spatial domains with data observed at irregularly spaced locations. Additionally, 
all these methods ignore the orthogonal constraints in MCA patterns.

In this paper, we propose a regularization approach of MCA that incorporates smoothness and localized features in dominant coupled patterns. The proposed method, called spatial MCA (abbreviated as SpatMCA), is applicable to data measured irregularly in space. 
In addition, the resulting estimates can be effectively computed using the alternating direction method of multipliers (ADMM) \citep{admm}.

The remainder of this paper is organized as follows. In Section~\ref{sec: proposal}, we introduce the proposed SpatMCA method,
including dominant coupled patterns estimation and spatial cross-covariance function estimation.
Our ADMM algorithm for computing the SpatMCA estimate is provided in Section~\ref{sec:algorithm}.
Numerical experiments that illustrate the superiority of SpatMCA and an application to study the relationship between sea surface temperature and precipitation datasets are presented in Section~\ref{sec:numerical}.

\section{The Proposed Method}
\label{sec: proposal}
Consider a sequence of uncorrelated, zero-mean, bivariate $L^2$-continuous spatial processes on spatial domains $D_1 \subset \mathbb{R}^d$ and $D_2 \subset \mathbb{R}^d$,
\[\{(\eta_{1i}(\bm{s}_1),\eta_{2i}(\bm{s}_2)) :\bm{s}_1\in D_1, \bm{s}_2\in D_2\};\quad  {i=1,\dots,n},\]
which have a common spatial covariance function $C_{jk}(\bm{s}_j,\bm{s}_k)=\mathrm{cov}(\eta_{ji}(\bm{s}_j),\eta_{ki}(\bm{s}_k))$; for $j,k=1,2$.
According to \cite{azaiez2015karhunen}, $C_{12}(\bm{s}_1,\bm{s}_2)$ can be decomposed as $C_{12}(\bm{s}_1,\bm{s}_2) = \sum_{k=1}^\infty d_ku_k(\bm{s}_1)v_k(\bm{s}_2)$, where  $\{d_k\}$ are nonnegative singular values with $d_1\geq d_2\geq\cdots$,  and $\{u_k(\cdot)\}$ and $\{v_k(\cdot)\}$ are the two corresponding sets of orthonormal basis functions. The decomposition is similar to the Karhunen-Lo\'{e}ve expansion \citep{karhunen,loeve} for a univariate spatial process. Suppose we observe data  $\bm{Y}_{ji} = (Y_{ji}(\bm{s}_{j1}),\dots,Y_{ji}(\bm{s}_{jp_j}))'$  with added noise $\bm{\epsilon}_{ji} \sim (\bm{0}, \sigma_j\bm{I})$ at the $p_j$ spatial locations $\bm{s}_{j1},\dots,\bm{s}_{jp_j}\in D_j$ for $j=1,2$, according to 
\begin{equation}
\left(\begin{array}{c}
\bm{Y}_{1i}\\ 
\bm{Y}_{2i}
\end{array}\right) =\left(\begin{array}{c}
\bm{\eta}_{1i}\\ 
\bm{\eta}_{2i}
\end{array}\right)+\left(\begin{array}{c}
\bm{\epsilon}_{1i}\\ 
\bm{\epsilon}_{2i}
\end{array}\right);\quad  {i=1,\dots,n},
\label{eq:measurement}
\end{equation}
where $\bm{\eta}_{ji} = (\eta_{ji}(\bm{s}_{j1}),\dots,\eta_{ji}(\bm{s}_{jp_j}))'$, and $\bm{\epsilon}_{1i}, 
\bm{\epsilon}_{2i}$ and $(\bm{\eta}_{1i},\bm{\eta}_{2i})$ are mutually uncorrelated. Assume $d_{K+1} = 0$, and denote the cross-covariance matrix between $\bm{\eta}_{1i}$ and $\bm{\eta}_{2i}$ by $\bm{\Sigma}_{12} =\mathrm{cov}(\bm{\eta}_{1i},\bm{\eta}_{2i})$. Let $\bm{\Sigma}_{12} = \bm{U} \bm{D}\bm{V}'$ be the SVD of $\bm{\Sigma}_{12}$, where  $\bm{D}=\mathrm{diag}(d_1,\dots,d_{K})$, $\bm{U} = {(}\bm{u}_1,\dots,\bm{u}_{K}) $ is a $ p_1\times K$  matrix with the $(k,{i})$-th element $u_k({\bm{s}_{1{i}}})$, and $\bm{V}=(\bm{v}_1,\dots,\bm{v}_{K})$ is a $p_2\times K$ matrix with the $(k,{i})$-th element $v_k({\bm{s}_{2{i}}})$. We aim to identify the first $L\leq K$ dominant spatial coupled patterns $(u_1(\cdot),\dots,u_{{L}}(\cdot))$ and $(v_1(\cdot),\dots,v_{L}(\cdot))$ with large $d_1,\dots,d_{L}$ for processes $\eta_1({\cdot})$ and $\eta_2({\cdot})$, as well as to estimate $C_{12}(\cdot,\cdot)$.

Let $\bm{Y}_j = (\bm{Y}_{j1}, \dots,\bm{Y}_{jn})'$ for $j=1,2$. The sample cross-covariance matrix of $\bm{Y}_1$ and $\bm{Y}_2$ is $\bm{S}_{12} = \bm{Y}'_1\bm{Y}_2/n$. Then the MCA estimates of $\bm{u}_k$ and $\bm{v}_k$ {obtained by the SVD of} $\bm{S}_{12}$ are $\tilde{\bm{u}}_k$ and $\tilde{\bm{v}}_k$, the $k$-th left and right singular vectors of $\bm{S}_{12}$, for $k=1,\dots,K$. Let $\tilde{\bm{U}} = (\tilde{\bm{u}}_1,\dots,\tilde{\bm{u}}_K)$ and $\tilde{\bm{V}} = (\tilde{\bm{v}}_1,\dots,\tilde{\bm{v}}_K)$ be  $p_1\times K$ and $p_2 \times K$ matrices formed by the first $K$ left and right singular vectors {of} $\bm{S}_{12}$. Then $(\tilde{\bm{U}},\tilde{\bm{V}})$ solves the following constrained optimization problem {\citep{svd_trace}}:
\[
\max_{\bm{U},\bm{V}} \mathrm{tr}(\bm{U}' \bm{S}_{12} \bm{V})\quad \mbox{subject to $\bm{U}'\bm{U}=\bm{V}'\bm{V}=\bm{I}_K$}, 
\]
where $\bm{U}=(\bm{u}_1,\dots,\bm{u}_K)$ and $\bm{V}=(\bm{v}_1,\dots,\bm{v}_K)$. However, $(\tilde{\bm{U}},\tilde{\bm{V}})$ may suffer from high estimation variability when $p_1$ or $p_2$ is large, $n$ is small, or $\sigma^2_1$ or $\sigma^2_2$ is large.
Consequently, the patterns of $(\tilde{\bm{U}},\tilde{\bm{V}})$ may be too noisy to be physically interpretable. 
Additionally, for continuous spatial domains $D_1$ and $D_2$, we also need to estimate $(u_k(\bm{s}^*_1), v_k(\bm{s}^*_2))$ at locations $\bm{s}^*_1{\in D_1}$ and $\bm{s}^*_2{\in D_2}$, where data may be {unavailable}.

\subsection{Regularized Spatial MCA}
To reduce high estimation variability of MCA while controlling bias, our main idea is to introduce some spatial structure. We propose a regularization approach by maximizing the following objective function:
\begin{eqnarray}
\mathrm{tr}(\bm{U}' \bm{S}_{12} \bm{V})- \sum_{k=1}^K \left\{{\tau_{1u}J(u_k)+\tau_{2u} \|\bm{u}_k\|_1+\tau_{1v}J(v_k) +\tau_{2v}\|\bm{v}_k\|_1}\right\},
\label{eq:obj1}
\end{eqnarray}	
over $u_1(\cdot),\dots,u_{K}(\cdot)$ and $v_1(\cdot),\dots,v_{K}(\cdot)$, subject to $\bm{U}'\bm{U}=\bm{V}'\bm{V}=\bm{I}_K$ and $\bm{u}'_1\bm{S}_{12}\bm{v}_1\geq\dots\geq\bm{u}'_K\bm{S}_{12}\bm{v}_K$, where
\[
J(u)=\sum_{z_1+\cdots+z_d=2}\int_{\mathcal{R}^d}\left(
\frac{\partial^2 u(\bm{s})}{\partial x_1^{z_1}\dots\partial x_d^{z_d}}\right)^2 d\bm{s},
\]			
is a roughness penalty, $\|\bm{u}_k\|_1 = \sum_{{i}=1}^{p_1}u_k(\bm{s}_{1{i}})$, $\|\bm{v}_k\|_1 = \sum_{{i}=1}^{p_2}v_k(\bm{s}_{2{i}})$, $\bm{s}=(x_1,\dots,x_d)'$, $\tau_{1u}$ and $\tau_{1v }$ are nonnegative smoothness parameters, and $\tau_{2u}$ and $\tau_{2v}$ are nonnegative sparseness parameters. Since the patterns of $u_k(\cdot)$ and $v_k(\cdot)$ could be very different, we allow $\tau_{1u} \neq \tau_{1v}$ and $\tau_{2u} \neq \tau_{2v}$. Note that $J(\cdot)$ is the smoothing spline penalty,  designed to enhance smoothness of $u_k(\cdot)$ and $v_k(\cdot)$, and the $L_1$ Lasso penalty {\citep{lasso}} is applied to seek sparse patterns by shrinking $u_k(\cdot)$ and $v_k(\cdot)$ toward zero. The combination of the smoothness and sparseness penalties was shown by \citet{spatpca} to be effective {in obtaining} smooth and localized patterns for a univariate spatial process. Denote $\hat{u}_1(\cdot),\dots,\hat{u}_K(\cdot)$ and $\hat{v}_1(\cdot),\dots,\hat{v}_K(\cdot)$ as the maximizers of \eqref{eq:obj1}. When $\tau_{1u}$ is larger, $\{\hat{u}_k(\cdot)\}$ become smoother, and {vice versa}. 
When $\tau_{2u}$ is larger, {$\{\hat{u}_k(\cdot)\}$} become more localized by forcing more elements of $\bm{u}_k$ to be zero. Similar results can be applied to $\tau_{1v}$ and $\tau_{2v}$ for {$\{\hat{v}_k(\cdot)\}$}. On the other hand, when $\tau_{1u}=\tau_{2u}=\tau_{1v}=\tau_{2v}=0$, the estimates reduce to the MCA estimates. 

According to the smoothing spline theory {\citep{nonparametric}}, $\hat{u}(\cdot)$ and $\hat{v}(\cdot)$ are natural cubic splines and thin-plate splines for $d=1$ and $d\in{\{2,3\}}$ {with knots at $\{\bm{s}_{11},\dots, \bm{s}_{1p_1}\}$ and $\{\bm{s}_{21},\dots, \bm{s}_{2p_2}\}$, respectively}. Specifically,
\begin{eqnarray}
\hat{u}_k(\bm{s}_1)&={\displaystyle\sum_{i=1}^{p_1}} {a}_{1i} g(\|\bm{s}_1-\bm{s}_{1i}\|)+b_{10}+{\displaystyle\sum_{j=1}^{d}} {b}_{{1j}} x_{1j},\label{eq:basis_u}\\
\hat{v}_k(\bm{s}_2)&={\displaystyle\sum_{i=1}^{p_2}} {a}_{2i} g(\|\bm{s}_2-\bm{s}_{2i}\|)+b_{20}+{\displaystyle\sum_{j=1}^{d}} {b}_{{2j}}x_{2j},\:
\label{eq:basis_v}
\end{eqnarray}

\noindent where $\bm{s}_{{j}}=(x_{{{j}}1},\dots,x_{{{j}}d})'$ for ${{j}}=1,2$,
\[
g(r) = \left\{
\begin{array}{ll}
\displaystyle\frac{1}{16\pi}r^{2}\log{r};  & \mbox{if $d=2$,}\smallskip\\
\displaystyle\frac{\Gamma(d/2-2)}{16\pi^{d/2}}r^{4-d}; & \mbox{if }d=1,3,\\
\end{array}\right.	 
\]
and the coefficients ${\bm{a}_j}=\left({a}_{j1},\dots,{a}_{jp_j}\right)'$
and ${\bm{b}_j}=\left({b}_{j0},b_{j1},\dots,{b}_{jd}\right)'$ for $j=1,2$ satisfy
\[
{\left(\begin{array}{cc}
	\bm{G}_1 & \bm{E}_1 \\
	\bm{E}'_1 & \bm{0} \\
	\end{array}\right) \left(\begin{array}{c}
	{\bm{a}_1}\\
	{\bm{b}_1}
	\end{array}\right)=\left(\begin{array}{c}
	\hat{\bm{u}}_k\\
	\bm{0}
	\end{array}\right)}
\quad
\mbox{and}\quad{
	\left(\begin{array}{cc}
	\bm{G}_2 & \bm{E}_2 \\
	\bm{E}'_2 & \bm{0} \\
	\end{array}\right) \left(\begin{array}{c}
	{\bm{a}_2}\\
	{\bm{b}_2}
	\end{array}\right)=\left(\begin{array}{c}
	\hat{\bm{v}}_k\\
	\bm{0}
	\end{array}\right){.}}
\]
Here $\hat{\bm{u}}_k = (\hat{u}_k(\bm{s}_{11}),\dots,\hat{u}_k(\bm{s}_{1p_1}))'$, $\hat{\bm{v}}_k = (\hat{v}_k(\bm{s}_{21}),\dots,\hat{v}_k(\bm{s}_{2p_2}))'$, $\bm{G}_j$ is a $p_j\times p_j$ matrix with the $(i,{i'})$-th element $g(\|\bm{s}_{{ji}}-\bm{s}_{j{i'}}\|)$,
and $\bm{E}_j$ is a $p_j\times (d+1)$ matrix with the $i$-th row $(1,\bm{s}'_{ji})$ for $j=1,2$.
Therefore, $\hat{u}_k(\cdot)$ and $\hat{v}_k(\cdot)$ in \eqref{eq:basis_u} and \eqref{eq:basis_v} can be expressed in terms of $\hat{\bm{u}}_k$ and $\hat{\bm{v}}_k${,} respectively.

The roughness penalties of $u_k(\cdot)$ and $v_k(\cdot)$ can also be written as
\begin{equation}
\label{eq:ch5smoothness}
J(u_k) =\bm{u}'_k \bm\Omega_1 \bm{u}_k \quad \mbox{and} \quad J(v_k) =\bm{v}'_k \bm\Omega_2 \bm{v}_k,
\end{equation}
where $\bm\Omega_j$ is a known $p_j\times p_j$ matrix determined only by $\bm{s}_{j1},\dots,\bm{s}_{jp_j}$ for $j=1,2$ {\citep{nonparametric}}. Therefore, from (\ref{eq:obj1}) and (\ref{eq:ch5smoothness}), the proposed estimate $(\hat{\bm{U}}_{K,\tau_{1u},\tau_{2u}}, \hat{\bm{V}}_{K,\tau_{1v},\tau_{2v}})$ of $(\bm{U},\bm{V})$ can be simplified by maximizing the following objective function:
\begin{eqnarray}
\label{eq:obj2}
\mathrm{tr}(\bm{U}' \bm{S}_{12} \bm{V})- \sum_{k=1}^K \left\{{\tau_{1u}\bm{u}'_k\bm\Omega_1\bm{u}_k+\tau_{2u} \|\bm{u}_k\|_1+ \tau_{1v}\bm{v}'_k\bm\Omega_2\bm{v}_k+ \tau_{2v} \|\bm{v}_k\|_1}\right\}, 
\end{eqnarray}	
subject to $\bm{U}'\bm{U}=\bm{V}'\bm{V}=\bm{I}_K$ and $\bm{u}'_1\bm{S}_{12}\bm{v}_1\geq\dots\geq\bm{u}'_K\bm{S}_{12}\bm{v}_K$. We call the proposed method based on \eqref{eq:obj2} SpatMCA. Given $(\hat{\bm{U}}_{K,\tau_{1u},\tau_{2u}}$,$\hat{\bm{V}}_{K,\tau_{1v},\tau_{2v}})$, the estimates of $(u_1(\cdot),v_1(\cdot)),\dots,(u_K(\cdot),v_K(\cdot))$ can be directly calculated by \eqref{eq:basis_u} and \eqref{eq:basis_v}.
Note that the SpatMCA estimate of (\ref{eq:obj2}) reduces to a sparse CCA estimate of  \cite{witten2009penalized} 
if {$\mathrm{var}(\bm{Y}_j)=\bm{I}_{p_j}$}, $\bm{\Omega}_j=\bm{I}_{p_j}$ for $j=1,2$, and the orthogonal constraints of $\bm{U}$ and $\bm{V}$ are dropped.
\subsection{Estimation of Cross-Covariance Function}

To estimate $C_{12}(\cdot,\cdot)$, we also have to estimate $\bm{D}$. Given $(\hat{\bm{U}},\hat{\bm{V}})=(\hat{\bm{U}}_{K,\tau_{1u},\tau_{2u}},\hat{\bm{V}}_{K,\tau_{1v},\tau_{2v}})$   with $\hat{\bm{U}} = (\hat{\bm{u}}_1,\dots,\hat{\bm{u}}_K)$ and $\hat{\bm{V}} = (\hat{\bm{v}}_1,\dots,\hat{\bm{v}}_K)$, the proposed estimate of $\bm{D}$ is 
\begin{equation}\label{eq:estimate_d}
\hat{\bm{D}} = \mathop{arg\min}_{d_1,\dots,d_K\geq 0}\|\bm{S}_{12} - \hat{\bm{U}}\bm{D}{\hat{\bm{V}}'}\|^2_F =\mathrm{diag}(\hat{d}_1,\dots,\hat{d}_K),
\end{equation}
where $\hat{d}_k = \max\{\hat{\bm{u}}'_k\bm{S}_{12}\hat{\bm{v}}_k,0\};$ $k=1,\dots,K$, and $\|\bm{M}\|_F=
\Big(\displaystyle\sum_{i,j}m^2_{ij}\Big)^{1/2}$ is the Frobenius norm of a matrix $\bm{M}$. Then, the proposed estimate of $C_{12}(\cdot,\cdot)$ is 
\begin{equation}\label{eq:estimate_crosscov}
\hat{C}_{12}(\bm{s}_1,\bm{s}_2)=\sum_{k=1}^K \hat{d}_k \hat{u}_k(\bm{s}_1)\hat{v}_k(\bm{s}_2).
\end{equation}
\subsection{Tuning Parameter Selection}
An $M$-fold cross-validation (CV) is applied to select the tuning parameters $\tau_{1u}$, $\tau_{2u}$, $\tau_{1v}$ and $\tau_{2v}$. First, we randomly decompose the index set $\{1,\dots,n\}$ into $M$ parts {that are} as close to the same size, $n_M$, as possible. Let $(\bm{Y}^{(m)}_1,\bm{Y}^{(m)}_2)$ be the sub-matrix of $(\bm{Y}_1,\bm{Y}_2)$ corresponding to the $m$-th part.
For $m=1,\dots, M$, we treat $(\bm{Y}^{(m)}_1,\bm{Y}^{(m)}_2)$  as the validation data,
and {we} obtain the estimate $(\hat{\bm{U}}^{(-m)}_{K,\tau_{1u},\tau_{2u}},\hat{\bm{V}}^{(-m)}_{K,\tau_{1v},\tau_{2v}})$ of $(\bm{U},\bm{V})$ for ${\{}\tau_{1u}, \tau_{2u},\tau_{1v},\tau_{2v}{\}}\in\mathcal{A}$
based on the remaining data  $(\bm{Y}^{(-m)}_1,\bm{Y}^{(-m)}_2)$ using the proposed method \eqref{eq:obj2}, where $\mathcal{A}\subset[0,\infty)^4$ is a candidate index set. Then the proposed CV criterion is
\begin{equation}
\label{eq:cv_criterion}
\mathrm{CV}({K,}\tau_{1u},\tau_{2u},\tau_{1v},\tau_{2v})=\frac{1}{M}\sum_{m=1}^M\|\bm{S}^{(m)}_{12} - \hat{\bm{U}}^{(-m)}_{K,\tau_{1u},\tau_{2u}}\hat{\bm{D}}^{(-m)}_{{K,\tau_{1u},\tau_{2u},\tau_{1v},\tau_{2v}}}(\hat{\bm{V}}^{(-m)}_{K,\tau_{1{v}},\tau_{2{v}}})'\|^2_F\:,
\end{equation}
where $\bm{S}^{(m)}_{12} = \left(\bm{X}^{(m)}\right)'\bm{Y}^{(m)}/n_M$, and $\hat{\bm{D}}^{(-m)}_{{K,\tau_{1u},\tau_{2u},\tau_{1v},\tau_{2v}}}$ is the estimate of $\bm{D}$ from \eqref{eq:estimate_d} with $({\hat{\bm{U}},\hat{\bm{V}}})$ replaced by $(\hat{\bm{U}}^{(-m)}_{K,\tau_{1u},\tau_{2u}},\hat{\bm{V}}^{(-m)}_{K,\tau_{1v},\tau_{2v}})$. 

Owing to the high computation cost to select {$\{\tau_{1u}, \tau_{2u},\tau_{1v},\tau_{2v}\}$} simultaneously {for each $K$}, 
we recommend an effective two-step procedure for selecting them. Specifically, we first select $\tau_{1u}$ and $\tau_{1v}$ with $\tau_{2u}=\tau_{2v}=0$ {by}
\begin{equation}\label{eq:cv_max1}
(\hat{\tau}_{1u}(K),\hat{\tau}_{1v}(K))=\displaystyle\mathop{\arg\min}_{\{\tau_{1u}, \tau_{1v}\}\subset[0,\infty)^2}\mathrm{CV}({K,}\tau_{1u},0,\tau_{1v},0),
\end{equation}
and then select $\tau_{2u}$ and $\tau_{2v}$ by
\begin{equation}\label{eq:cv_max2}
(\hat{\tau}_{2u}(K),\hat{\tau}_{2v}(K))=\displaystyle\mathop{\arg\min}_{\{\tau_{2u}, \tau_{2v}\}\subset[0,\infty)^2}\mathrm{CV}(K,\hat{\tau}_{1u}(K),\tau_{2u},\hat{\tau}_{1v}(K),\tau_{2v}).
\end{equation}

{Finally, we select the }rank $K$ of $\bm{U}\bm{D}\bm{V}'$ by computing the CV values of \eqref{eq:cv_criterion} for $K=1,2,\dots$, evaluated at the {four} selected tuning {parameter values} until no further reduction of the CV value is obtained. That is,
\begin{align}
\hat{K}=\min\{&K: \mathrm{CV}\left(K,\hat{\tau}_{1u}(K),\hat{\tau}_{2u}(K),\hat{\tau}_{1v}(K),\hat{\tau}_{2v}(K)\right)\leq\notag\\
~&\mathrm{CV}\left(K+1,\hat{\tau}_{1u}(K+1),\hat{\tau}_{2u}(K+1),\hat{\tau}_{1v}(K+1),\hat{\tau}_{2v}(K+1)\right);K=1,2,\dots\}.
\label{eq:ch5khat}
\end{align}

\section{Computation Algorithm}
\label{sec:algorithm}
Let $\bm{G} = (\bm{U}', \bm{V}')'$ be a $(p_1+ p_2)\times K$ matrix with the $({i},k)$-th element $g_{{i}k}$. The objective function \eqref{eq:obj2} can be rewritten as
\begin{equation}\label{eq:obj3}
\mathrm{tr}( \bm{G}'\bm{\Theta}\bm{G})-\sum_{k=1}^K \left(\tau_{2u} \sum_{{i}=1}^{p_1}|{g}_{{i}k}|+\tau_{2v} \sum_{{i}=p_1+1}^{p_1+p_2}|g_{{i}k}|\right),
\end{equation}
subject to $\bm{U}'\bm{U} = \bm{V}'\bm{V}=\bm{I}_{K}$, where $\bm{\Theta} ={\left(\begin{array}{cc}
	-\tau_{1u}\bm{\Omega}_{1} & \bm{S}_{12}/2 \\
	\bm{S}'_{12}/2 & -\tau_{1v}\bm{\Omega}_{2} \\
	\end{array}\right)}$.  The maximizer of \eqref{eq:obj3}, consisting of the orthogonal constraint and the Lasso penalty, is too complex to solve directly. We adopt the ADMM algorithm \citep[originated by][]{admm_2} by decomposing the constrained optimization problems into small subproblems that can be efficiently handled. The readers are referred to \citet{admm} for more details regarding ADMM.

First, we transform \eqref{eq:obj3} into the following equivalent {form} by adding $(p_1+p_2) \times K$ parameter matrices $\bm{Q}$ and  $\bm{R}${:}
\begin{equation*}\label{admm_opt}
\mathrm{tr}( \bm{G}'\bm{\Theta}\bm{G})-\sum_{k=1}^K \left(\tau_{2u} \sum_{i=1}^{p_1}|{r}_{ik}|+\tau_{2v} \sum_{i=p_1+1}^{p_1+p_2}|r_{ik}|\right),
\end{equation*}
subject to $\bm{Q}_1'\bm{Q}_1 = \bm{Q}_2'\bm{Q}_2=\bm{I}_{K}$, and a new constraint $\bm{G}=\bm{Q}=\bm{R}$, where $r_{ik}$ is the $(i,k)$-th element of $\bm{R}$, $\bm{Q}=(\bm{Q}'_1,\bm{Q}'_2)'$, and $\bm{Q}_1$ and $\bm{Q}_2$ {are} $p_1\times K$ and $p_2\times K$ sub-matrices of $\bm{Q}$ formed by the first $p_1$ and the last $p_2$ rows of $\bm{Q}$, respectively. The resulting augmented Lagrange function is		
\begin{align*}
L(\bm{G}, \bm{R}, \bm{Q}, \bm{\Gamma}_{1},\bm{\Gamma}_{2})=&~\mathrm{tr}( \bm{G}'\bm{\Theta}\bm{G})-\sum_{k=1}^K \left(\tau_{2u} \sum_{{i}=1}^{p_1}|{r}_{{i}k}|+\tau_{2v} \sum_{{i}=p_1+1}^{p_1+p_2}|{r}_{{i}k}|\right)\\
&~-\mathrm{tr}(\bm{\Gamma}'_{1}(\bm{G}-\bm{R}))-\mathrm{tr}(\bm{\Gamma}'_{2}(\bm{G}-\bm{Q}))\\&~-\frac{\zeta}{2}(\|\bm{G}-\bm{R}\|^2_F+\|\bm{G}-\bm{Q}\|^2_F),
\end{align*}subject to $\bm{Q}_1'\bm{Q}_1 = \bm{Q}_2'\bm{Q}_2=\bm{I}_{K}$, where $\bm{\Gamma}_{1}$ and $\bm{\Gamma}_{2}$ are $(p_1+p_2)\times K$ matrices of Lagrange multipliers, and $\zeta \geq 0$ is a penalty parameter to promote convergence. Then the ADMM steps at the $(\ell+1)$-th iteration {have the following} closed formed expressions:
\begin{align}
\bm{G}^{(\ell+1)}
=&~ \mathop{\arg\max}_{\bm{G}}L(\bm{G}, \bm{R}^{(\ell)}, \bm{Q}^{(\ell)}, \bm{\Gamma}_{1}^{(\ell)},\bm{\Gamma}_{2}^{(\ell)})
\notag\\
=&~ \frac{1}{2}(\zeta\bm{I}-\bm{\Theta})^{{-}1}\left(\zeta(\bm{R}^{(\ell)}{+}\bm{Q}^{(\ell)}){-}\bm{\Gamma}^{(\ell)}_{1}{-}\bm{\Gamma}^{(\ell)}_{2}\right),\label{eq:G}\\
\bm{R}^{(\ell+1)}
=&~\mathop{\arg\max}_{\bm{R}}L(\bm{G}^{(\ell+1)}, \bm{R}, \bm{Q}^{(\ell)}, \bm{\Gamma}_{1}^{(\ell)},\bm{\Gamma}_{2}^{(\ell)})
\notag\\
=&~ \left(\frac{1}{\zeta}\mathcal{S}_{\tau_{2}}\left(\zeta{g}^{(\ell+1)}_{{i}k}+{\gamma}_{1{i}k}^{(\ell)}\right)\right)_{(p_1+p_2)\times K}, \label{eq:ch5R}\\
\bm{Q}^{(\ell+1)}
=&~ \mathop{\arg\max}_{\bm{Q:\bm{Q}'\bm{Q}=\bm{I}}}L(\bm{G}^{(\ell+1)}, \bm{R}^{(\ell+1)}, \bm{Q}, \bm{\Gamma}_{1}^{(\ell)},\bm{\Gamma}_{2}^{(\ell)})\notag\\
=&~\left(\bm{F}_1^{(\ell)}\left(\bm{E}_1^{(\ell)}\right)',\bm{F}_2^{(\ell)}\left(\bm{E}_2^{(\ell)}\right)'\right)' ,\label{eq:ch5Q}\\
\bm{\Gamma}^{(\ell+1)}_{1}
=&~ \bm{\Gamma}^{(\ell)}_{1}+ \zeta\left(\bm{G}^{(\ell+1)}-\bm{Q}^{(\ell+1)}\right), \label{eq:ch5Gamma1}\\
\bm{\Gamma}^{(\ell+1)}_{2}
=&~ \bm{\Gamma}^{(\ell) }_{2}+ \zeta\left(\bm{G}^{(\ell+1)}-\bm{R}^{(\ell+1)}\right),\label{eq:ch5Gamma2}
\end{align}
where
\[\displaystyle \mathcal{S}_{\tau_{2}}(\gamma_{1jk})= \left\{
\begin{array}{ll}
\mathrm{sign}(\gamma_{1{i}k})\max(|\gamma_{1{i}k}|-\tau_{2u},0);  & \mbox{if ${i} \leq p_1$,}\smallskip\\
\mathrm{sign}(\gamma_{1{i}k})\max(|\gamma_{1{i}k}|-\tau_{2v},0); & \mbox{{if $i > p_1$}},\\
\end{array}\right.\]
$\gamma_{1{i}k}$ is the $({i},k)$-th element of $\bm{\Gamma}_1$, $\bm{E}_j^{(\ell)}\bm{\Lambda}_j^{(\ell)}\left(\bm{F}_j^{(\ell)}\right)'$ is the SVD of $\zeta\bm{G}_j^{(\ell+1)}+ \bm{\Gamma}^{(\ell)}_{2j}$ for $j=1,2$,  $\bm{G}^{(\ell+1)}_1$ and $\bm{G}^{(\ell+1)}_{21}$ are $p_1\times K$ and $p_2\times K$ sub-matrices of $\bm{G}^{(\ell+1)}$ corresponding to $\bm{U}$ and $\bm{V}$, and  $\bm{\Gamma}^{(\ell+1)}_{21}$ and $\bm{\Gamma}^{(\ell+1)}_{22}$ are $p_1\times K$ and $p_2\times K$ sub-matrices of $\bm{\Gamma}^{(\ell+1)}_2$ corresponding to $\bm{U}$ and $\bm{V}$. Note that $\zeta$ must be chosen large enough to ensure that $\zeta\bm{I}-\bm{\Theta}$ in \eqref{eq:G}
is positive-definite.

\section{Numerical Examples}
\label{sec:numerical}
This section contains several simulation examples in one-dimensional and two-dimensional spatial domains and an application of SpatMCA to a real dataset. We compared the performance of the proposed SpatMCA with three other methods: (1) MCA ($\tau_{1u}=\tau_{1v}=\tau_{2u}=\tau_{2v}=0$);
(2) SpatMCA with the smoothness penalties only ($\tau_{2u}=\tau_{2v}=0$);
(3) SpatMCA with the sparseness penalties only ($\tau_{1u}=\tau_{1v}=0$),
in terms of the following loss function:
\begin{equation}
\label{eq:loss1} 
\mathrm{Loss}(\hat{C}_{12}) =\frac{1}{p_1p_2} \sum_{i=1}^{p_1}\sum_{j=1}^{p_2}\big(\hat{C}_{12}(\bm{s}_{1i},\bm{s}_{2j})-C_{12}(\bm{s}_{1i},\bm{s}_{2j})\big)^2\:.
\end{equation}
Throughout this section, we applied the proposed SpatMCA method and the ADMM algorithm given by \eqref{eq:G}–\eqref{eq:ch5Gamma2} to compute the SpatMCA estimates
with $\zeta$ being ten times the maximum singular value of $\bm{S}_{12}$. Additionally, the stopping criterion for the ADMM algorithm is
\[
\frac{1}{\sqrt{p_1p_2}} \max\left( \|\bm{G}^{(\ell+1)}-\bm{G}^{(\ell)}\|_F,\|\bm{G}^{(\ell+1)}-\bm{R}^{(\ell+1)}\|_F,
\|\bm{G}^{(\ell+1)}-\bm{Q}^{(\ell+1)}\|_F \right)\leq 10^{-4}\:.
\]
\subsection{A One-Dimensional Experiment}\label{sec:1d}
We generated data from \eqref{eq:measurement} with $K=2$, {$d=1$, $n=1000,$}
\[\left(\begin{array}{c}
\bm{\eta}_{1i} \\
\bm{\eta}_{2i} \\
\end{array}\right)\sim N\left(\bm{0}, {\left(\begin{array}{cc}
	\bm{I} & \bm{U}\mathrm{diag}(d_1,d_2)\bm{V}' \\
	\bm{V}\mathrm{diag}(d_1,d_2)\bm{U}' & \bm{I}\\
	\end{array}\right)}\right),\] 
$\bm{\epsilon}_{ji}\sim N(\bm{0},\bm{I})$, ${p_j}=50$, $(\bm{s}_{j1},  \dots,\bm{s}_{jp_j})$ equally spaced in $[-7,7]$, and
\begin{align}
u_1(\bm{s}_1)
=&~ \frac{1}{c_1}\exp(-(x_{11}^2+\cdots+x_{1d}^2)),
\label{eq:u1_sim}\\
v_1(\bm{s}_2)
=&~ \frac{1}{c_2}\exp(-((x_{21}-2)^2+\cdots+(x_{2d}-2)^2)/2),
\label{eq:v1_sim}\\
u_2(\bm{s}_1)
=&~ \frac{1}{c_3}x_{11}\cdots x_{1d}\exp(-(x_1^2+\cdots+x_{1d}^2)),
\label{eq:u2_sim}\\
v_2(\bm{s}_2)
=&~ \frac{1}{c_4}(x_{21}-2)\cdots (x_{2d}-2)\exp(-((x_{21}-2)^2+\cdots+(x_{2d}-2)^2)/2),
\label{eq:v2_sim}
\end{align}
where $\bm{s}_j=(x_{j1},\dots,x_{jd})'$, $c_1$, $c_2$, $c_3$ and $c_4$ are normalization constants such that
$\|\bm{u}_j\|_2=\|\bm{v}_j\|_2=1$ for $j=1,2$. We considered three pairs of $(d_1,d_2)\in\{(1,0), (0.5,0), (1,0.7)\}$, and applied the proposed SpatMCA with $K=\{1,2,5\}$ and $\hat{K}$ selected by \eqref{eq:ch5khat}. For each case, we applied the 5-fold CV of \eqref{eq:ch5khat} to select ${\{}\tau_{1u}, \tau_{1v}, \tau_{2u}, \tau_{2v}{\}}$ among $21$ values of $\tau_{1u}$ and $\tau_{1v}$  (including $0$ and the other 20 values equally spaced on the log scale from $10^{-2}$ to $10$) and $11$ values of $\tau_{2u}$ and $\tau_{2v}$ (including $0$ and the other 10 values equally spaced on the log scale from $10^{-3}$ to $1$).

\begin{figure}\centering
	$\hat{u}_1(\cdot)$ based on $K=1$ for $(d_1,d_2)=(1,0)$\\
	\includegraphics[scale=0.39]{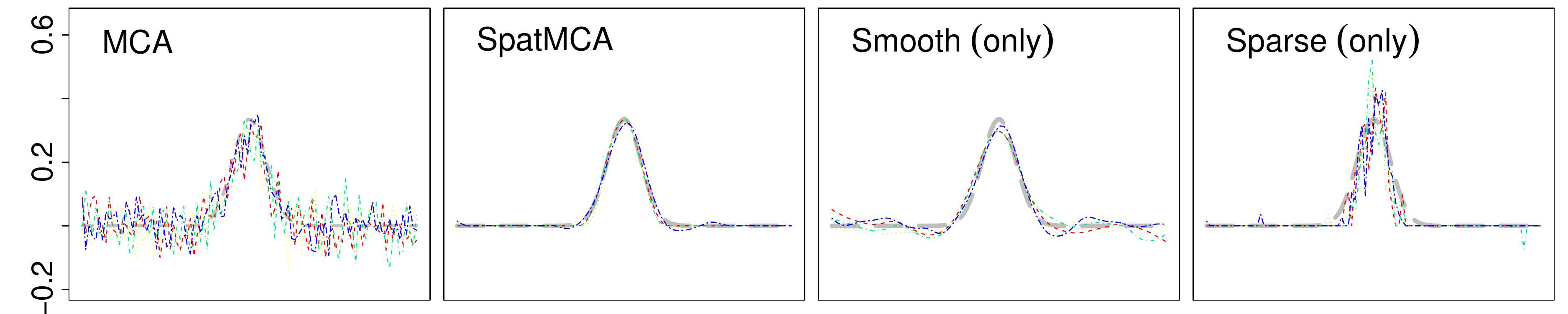}
	$\hat{u}_1(\cdot)$ based on $K=1$ for $(d_1,d_2)=(0.5,0)$\\
	\includegraphics[scale=0.39]{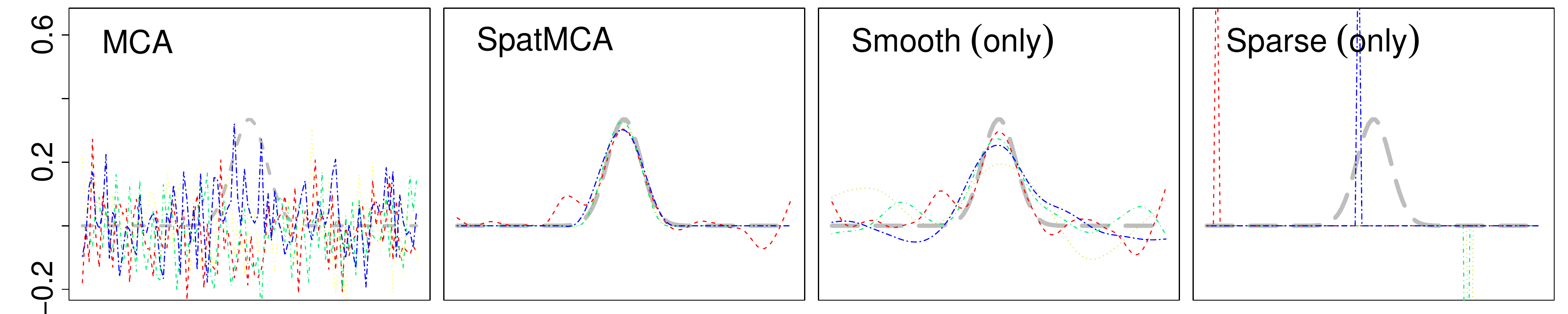}
	$\hat{u}_1(\cdot)$ based on $K=2$ for $(d_1,d_2)=(1,0.7)$\\
	\includegraphics[scale=0.39]{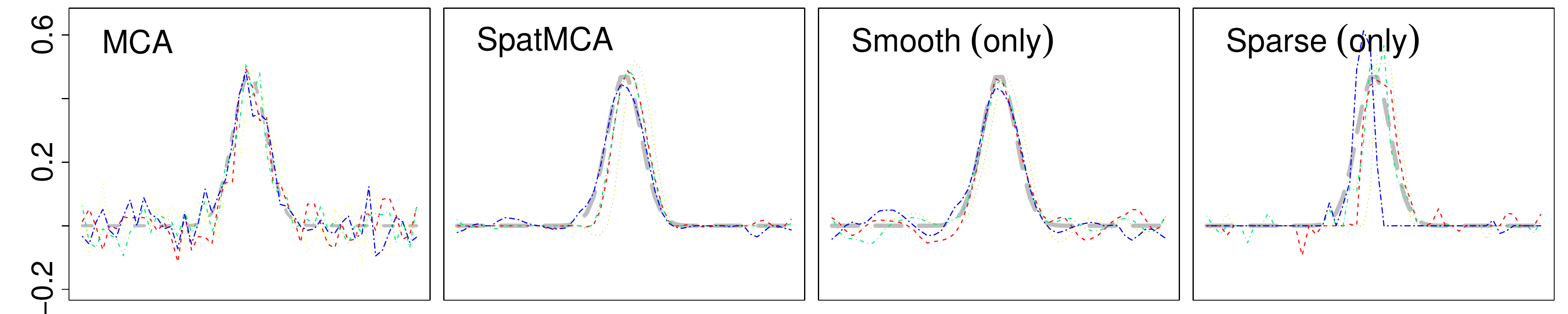}
	$\hat{u}_2(\cdot)$ based on $K=2$ for $(d_1,d_2)=(1,0.7)$\\
	\includegraphics[scale=0.39]{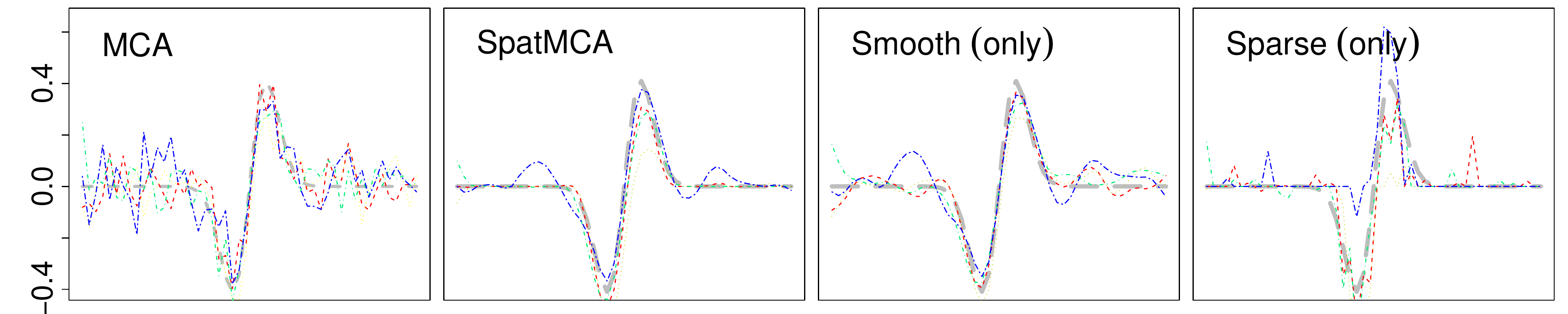}
	\caption{Estimates of $u_1(\cdot)$ and $u_2(\cdot)$ obtained from various methods based on data generated from three different combinations of singular values. Each panel consists of four estimates (in four different line types) corresponding to four randomly generated datasets, where the dash gray lines are the true $u_1(\cdot)$ and $u_2(\cdot)$.}
	\label{fig:est_u_d1}
\end{figure}

\begin{figure}\centering
	$\hat{v}_1(\cdot)$ based on $K=1$ for $ (d_1,d_2)=(1,0)$\\
	\includegraphics[scale=0.39]{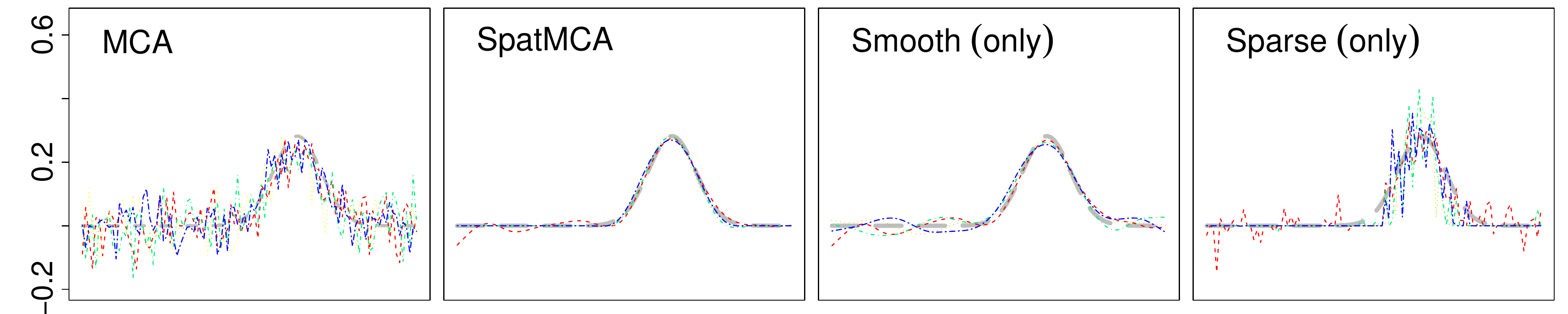}\\
	$\hat{v}_1(\cdot)$ based on $K=1$ for $ (d_1,d_2)=(0.5,0)$\\
	\includegraphics[scale=0.39]{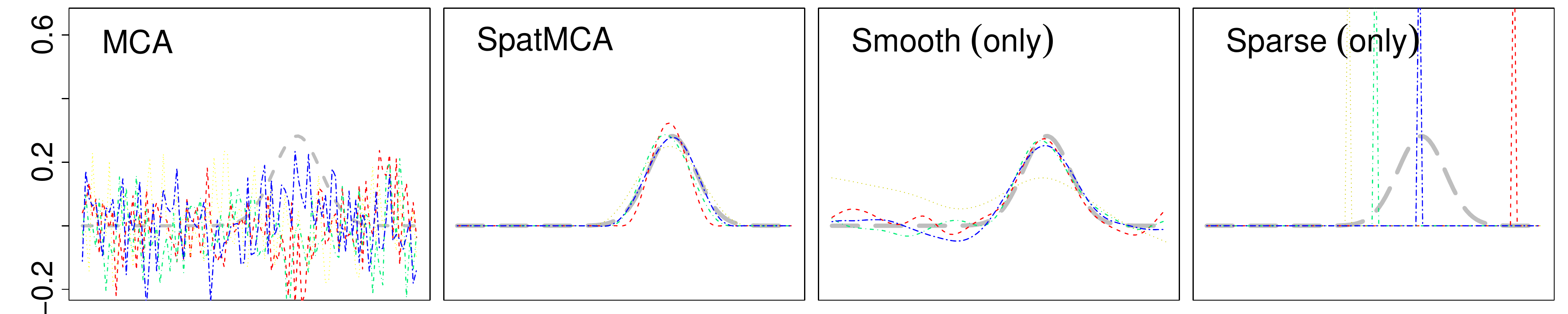}\\
	$\hat{v}_1(\cdot)$ based on $K=2$ for $ (d_1,d_2)=(1,0.7)$\\
	\includegraphics[scale=0.39]{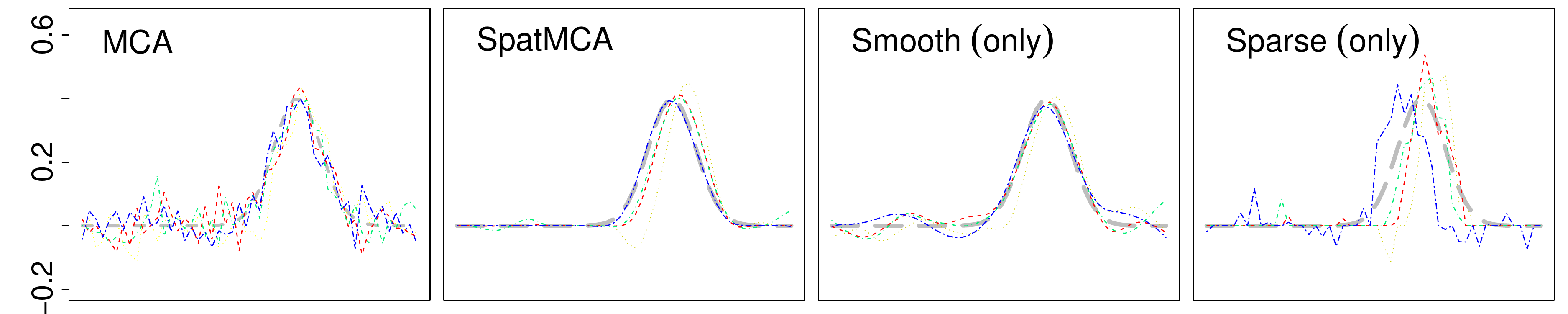}\\
	$\hat{v}_2(\cdot)$ based on $K=2$ for $  (d_1,d_2)=(1,0.7)$\\
	\includegraphics[scale=0.39]{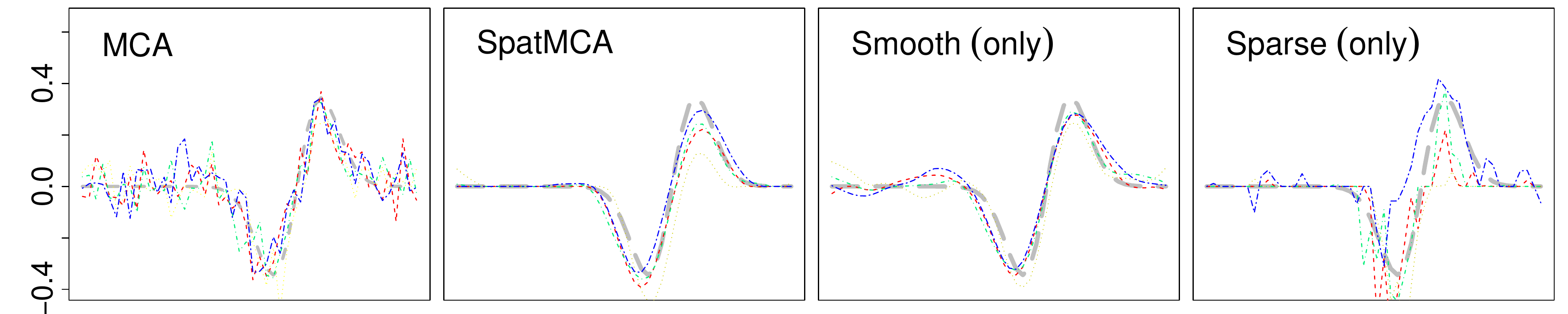}
	\caption{Estimates of $v_1(\cdot)$ and $v_2(\cdot)$ obtained from various methods based on data generated from three different combinations of singular values. Each panel consists of four estimates (in four different line types) corresponding to four randomly generated datasets, where the dash gray lines are the true $v_1(\cdot)$ and $v_2(\cdot)$.}
	\label{fig:est_v_d1}
\end{figure}

Figures~\ref{fig:est_u_d1} and \ref{fig:est_v_d1} show the estimates of $u_k(\cdot)$ and $v_k(\cdot)$, respectively, for the four methods based on three different combinations of singular values. Each case contains four estimated functions based on four randomly generated datasets.
Not surprisingly, the MCA estimates considering no spatial structure are very noisy, particularly when the signal-to-noise ratio is small.
Adding only the smoothness penalties (i.e., $\tau_{2u}=\tau_{2v}=0$) reduces noise, but introduces some bias. On the other hand, adding only the sparseness penalties (i.e, $\tau_{1u}=\tau_{1v}=0$) does not reduce much noise, despite that the estimated $\{u_k(\cdot)\}$ and $\{v_k(\cdot)\}$ are forced to be zeros at some locations. Our SpatMCA estimates generally reproduce the targets with little noise for all cases even for the small signal-to-noise ratio, indicating the effectiveness of regularization.

\begin{figure}\centering
	$(d_1,d_2)=(1,0)$
	\includegraphics[scale=0.42]{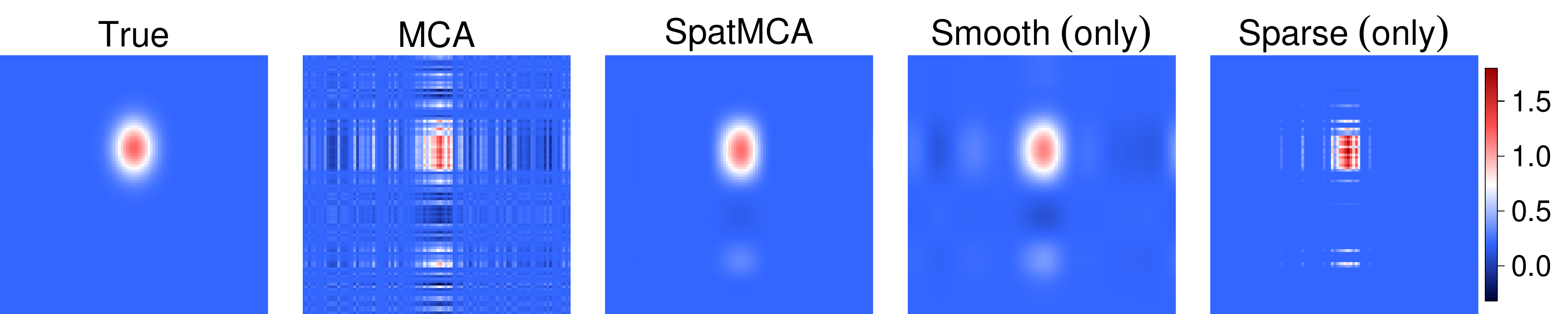}
	$(d_1,d_2)=(0.5)$
	\includegraphics[scale=0.42]{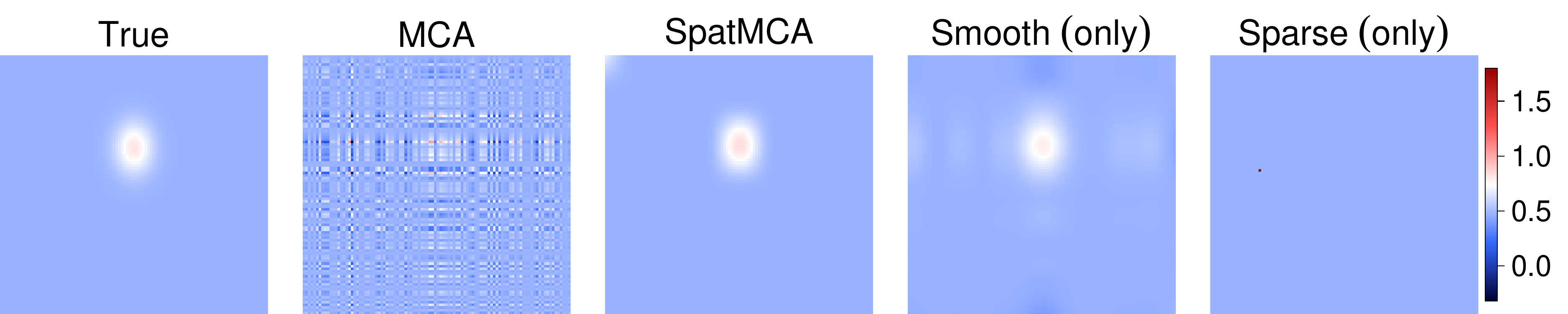}
	$(d_1,d_2)=(1,0.7)$
	\includegraphics[scale=0.42]{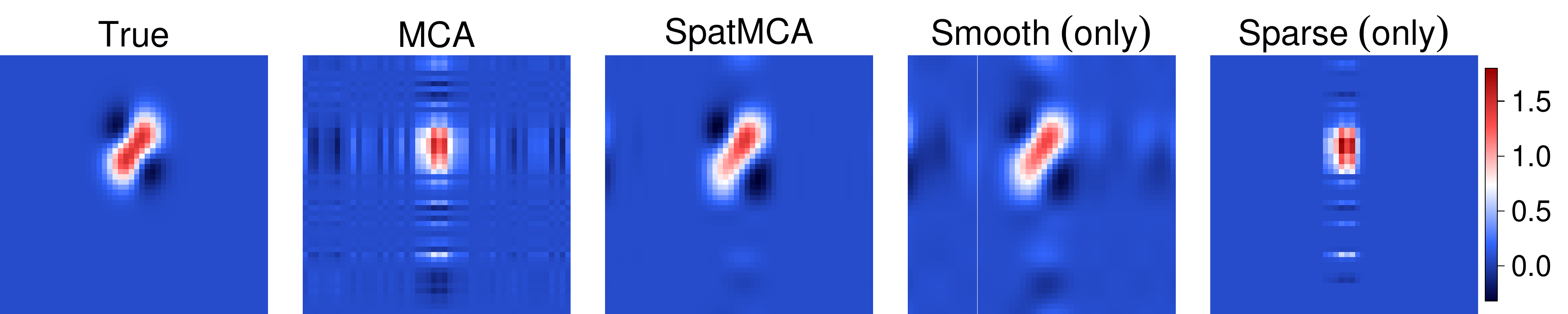}
	\caption{True cross-covariance functions and their estimates obtained from various methods with the rank $\hat{K}$ selected by CV for three different combinations of {singular values}.}
	\label{fig:cov_d1}
\end{figure}

\begin{figure}
	\begin{tabular}{ccc}
		{{$(d_1,d_2)=(1,0)$}, $K=1$}&{{$(d_1,d_2)=(0.5,0)$}, $K=1$}&{{$(d_1,d_2)=(1,0.7)$}, $K=1$}\\
		\includegraphics[scale=0.12]{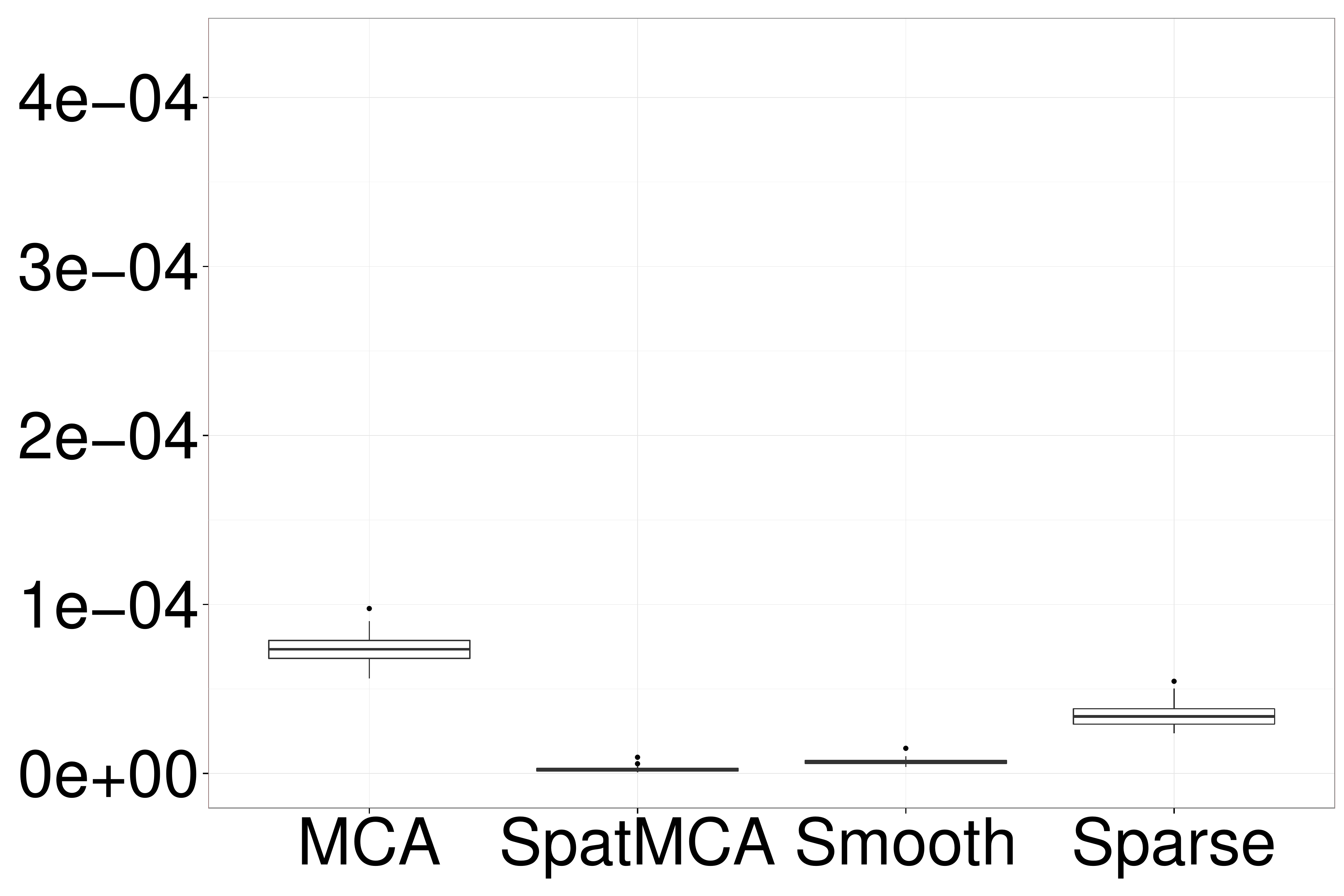}\hspace{4pt}&
		\includegraphics[scale=0.12]{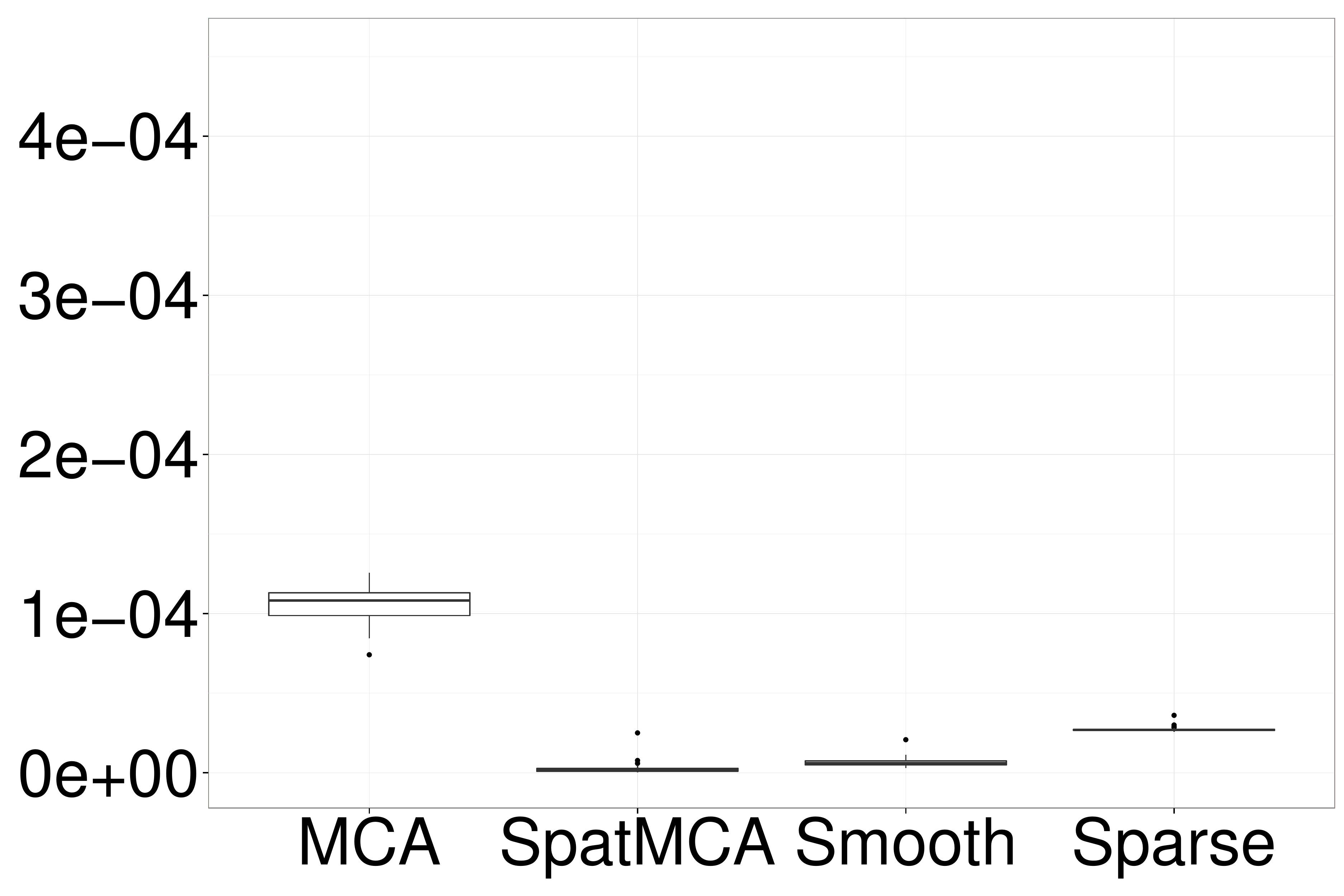}\hspace{4pt}&
		\includegraphics[scale=0.12]{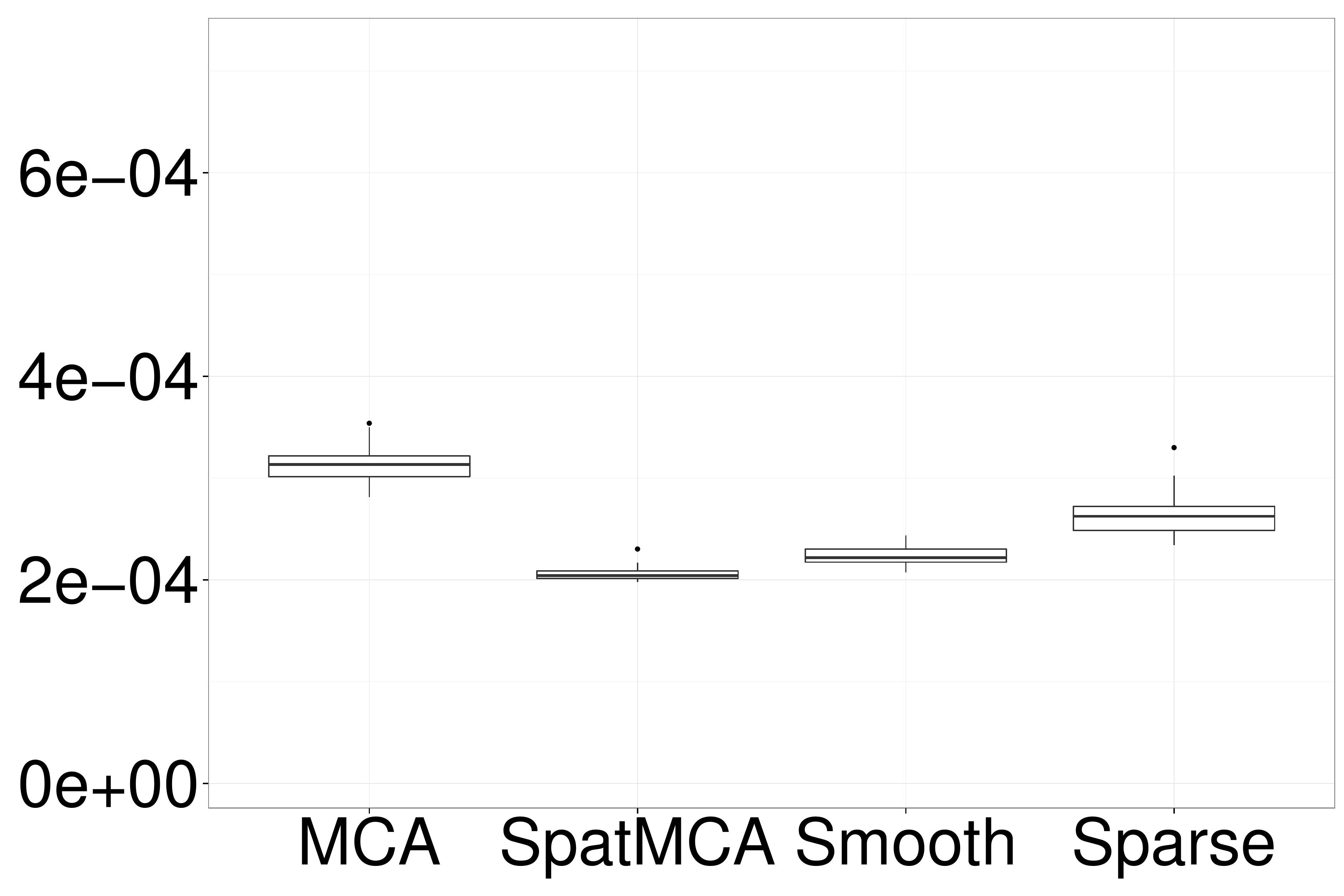}\hspace{4pt}\\
		{{$(d_1,d_2)=(1,0)$}, $K=2$}&{{$(d_1,d_2)=(0.5,0)$}, $K=2$}&{{$(d_1,d_2)=(1,0.7)$}, $K=2$}\\
		\includegraphics[scale=0.12]{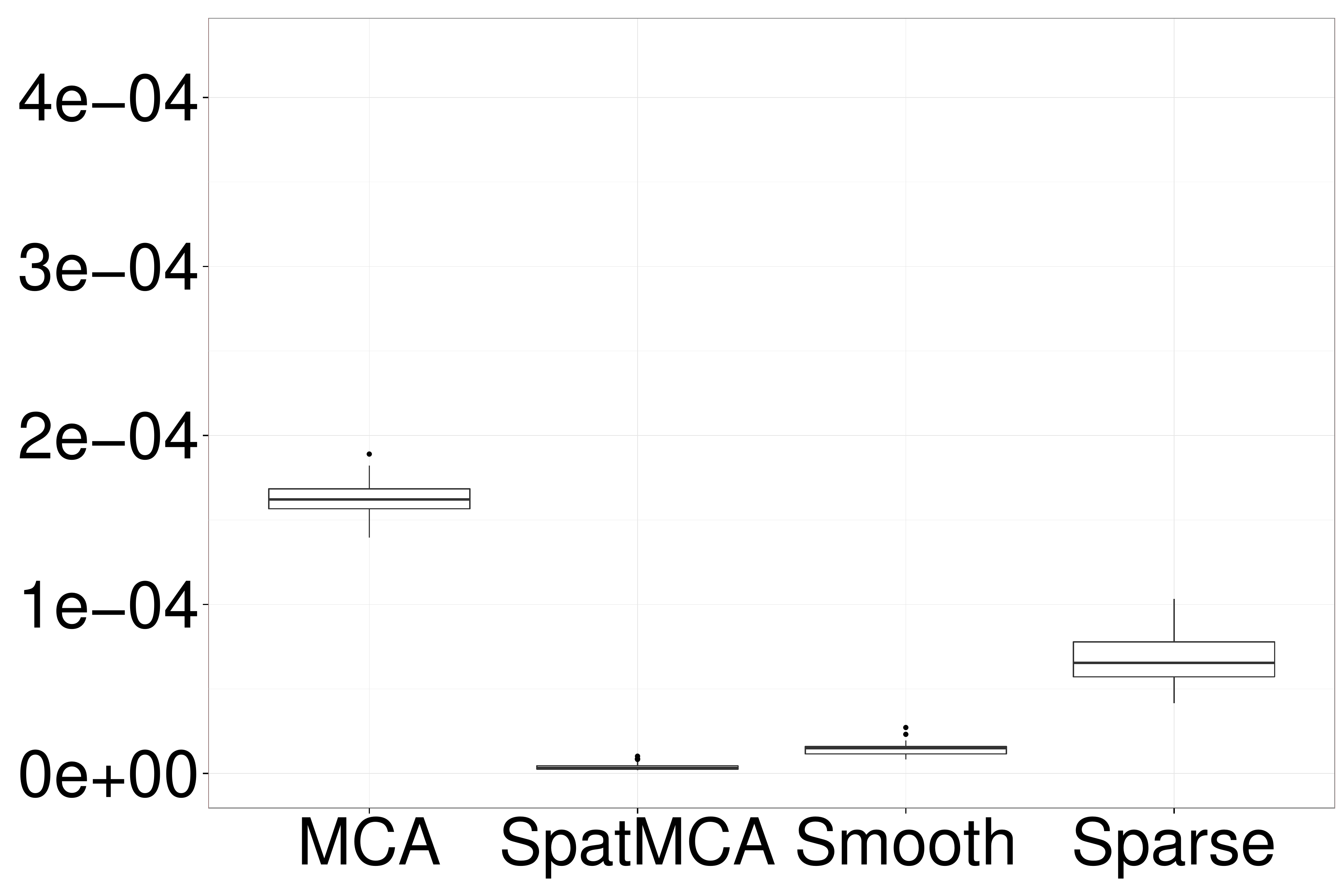}\hspace{4pt}&
		\includegraphics[scale=0.12]{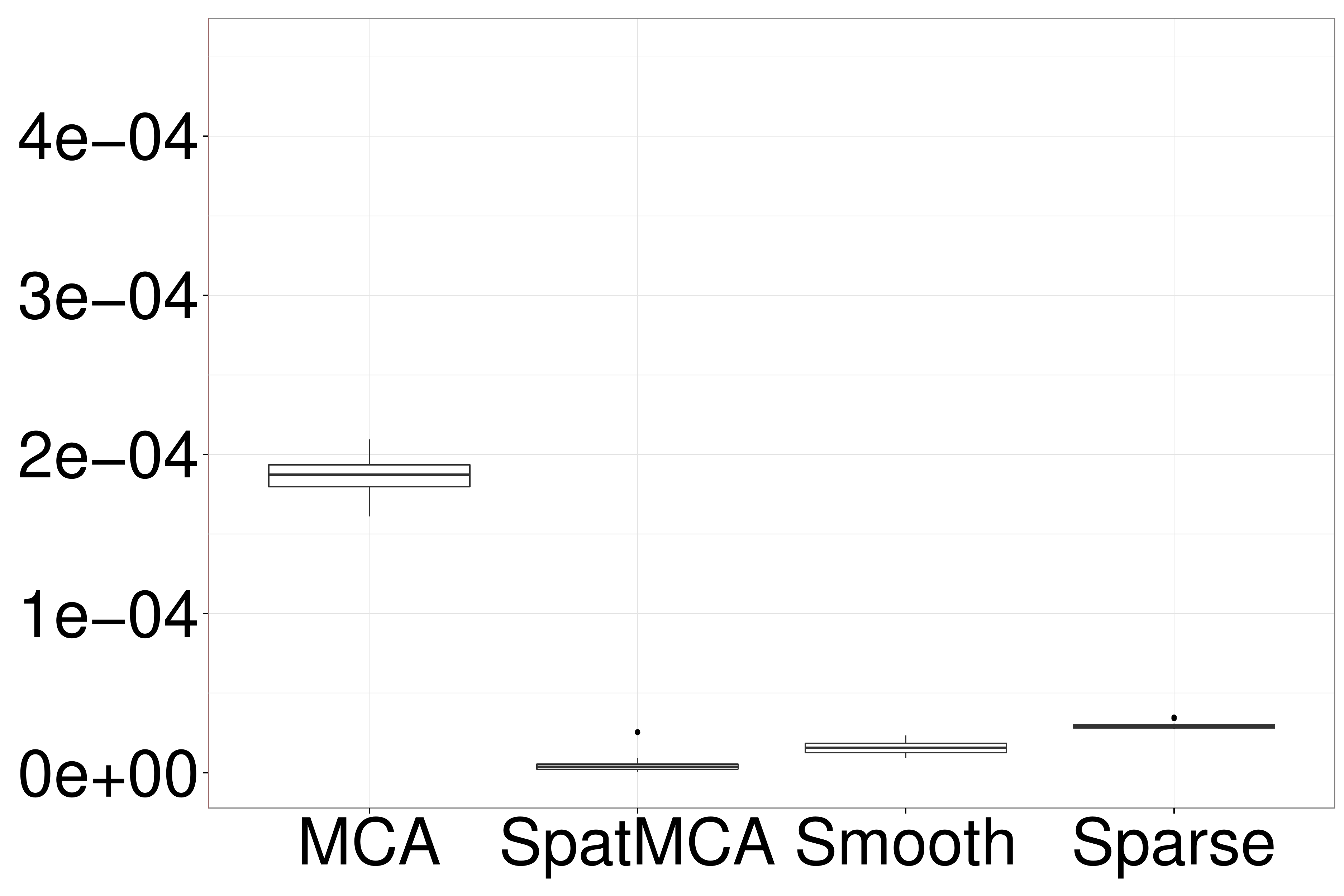}\hspace{4pt}&
		\includegraphics[scale=0.12]{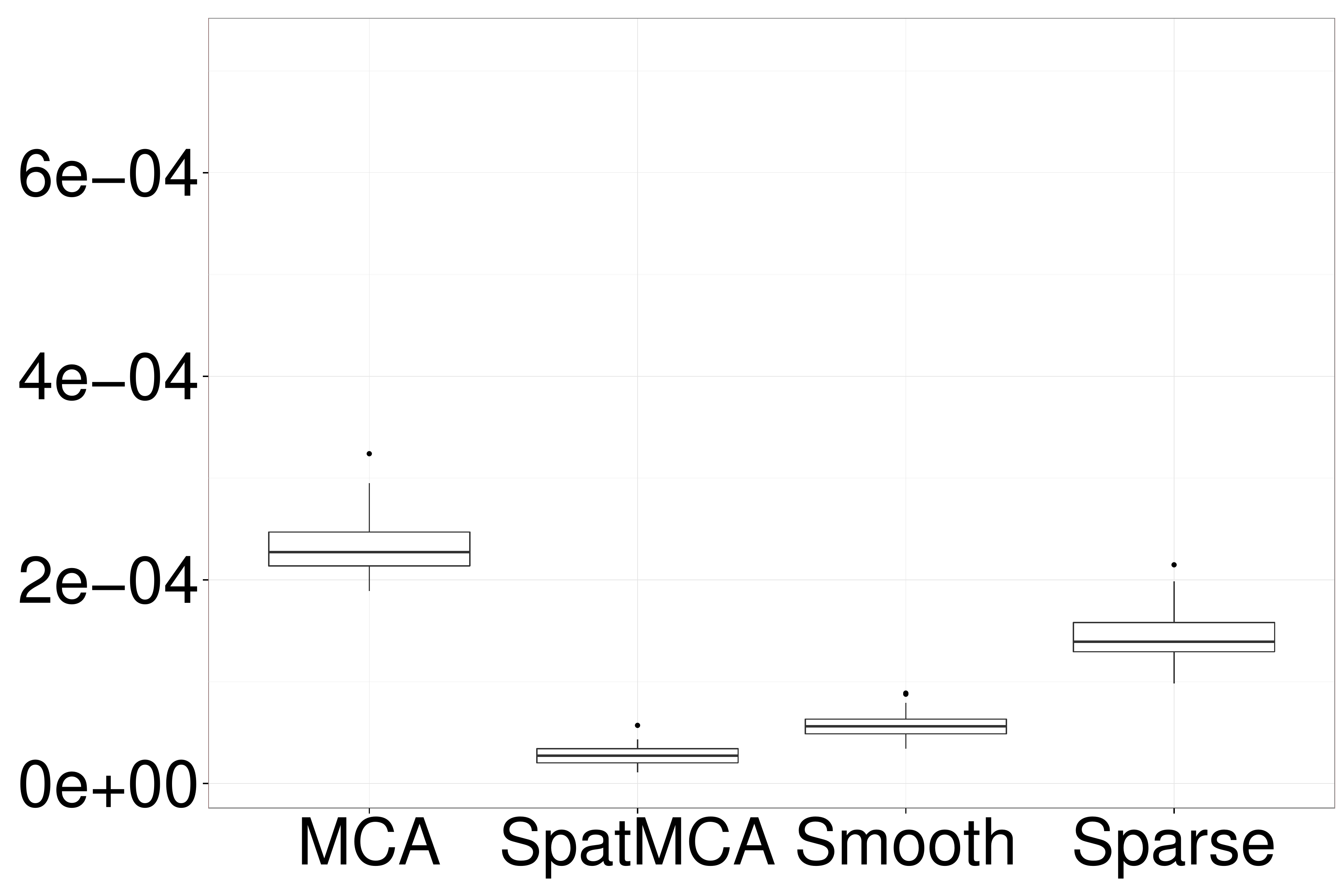}\hspace{4pt}\\
		{{$(d_1,d_2)=(1,0)$}, $K=5$}&{{$(d_1,d_2)=(0.5,0)$}, $K=5$}&{{$(d_1,d_2)=(1,0.7)$}, $K=5$}\\
		\includegraphics[scale=0.12]{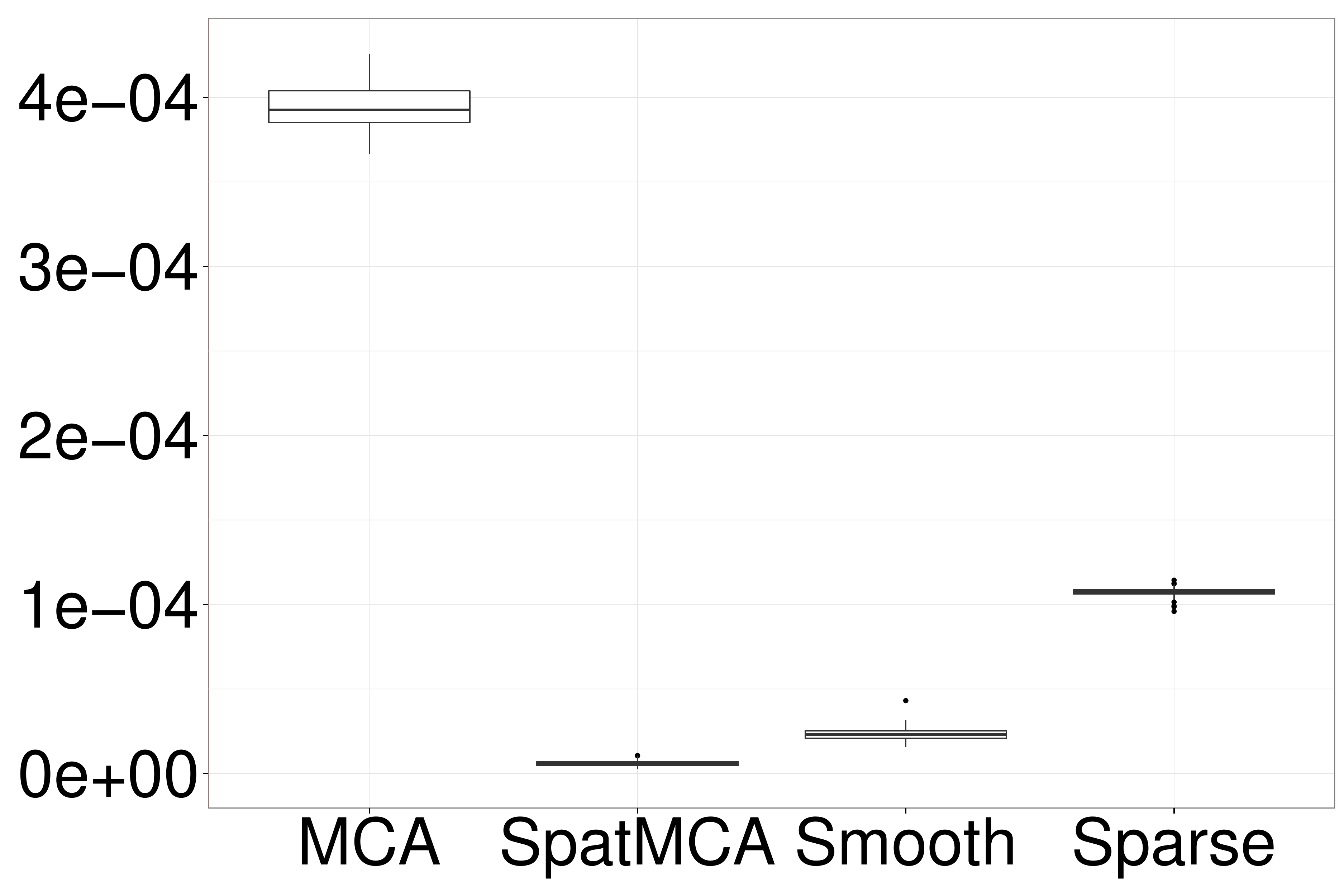}\hspace{4pt}&
		\includegraphics[scale=0.12]{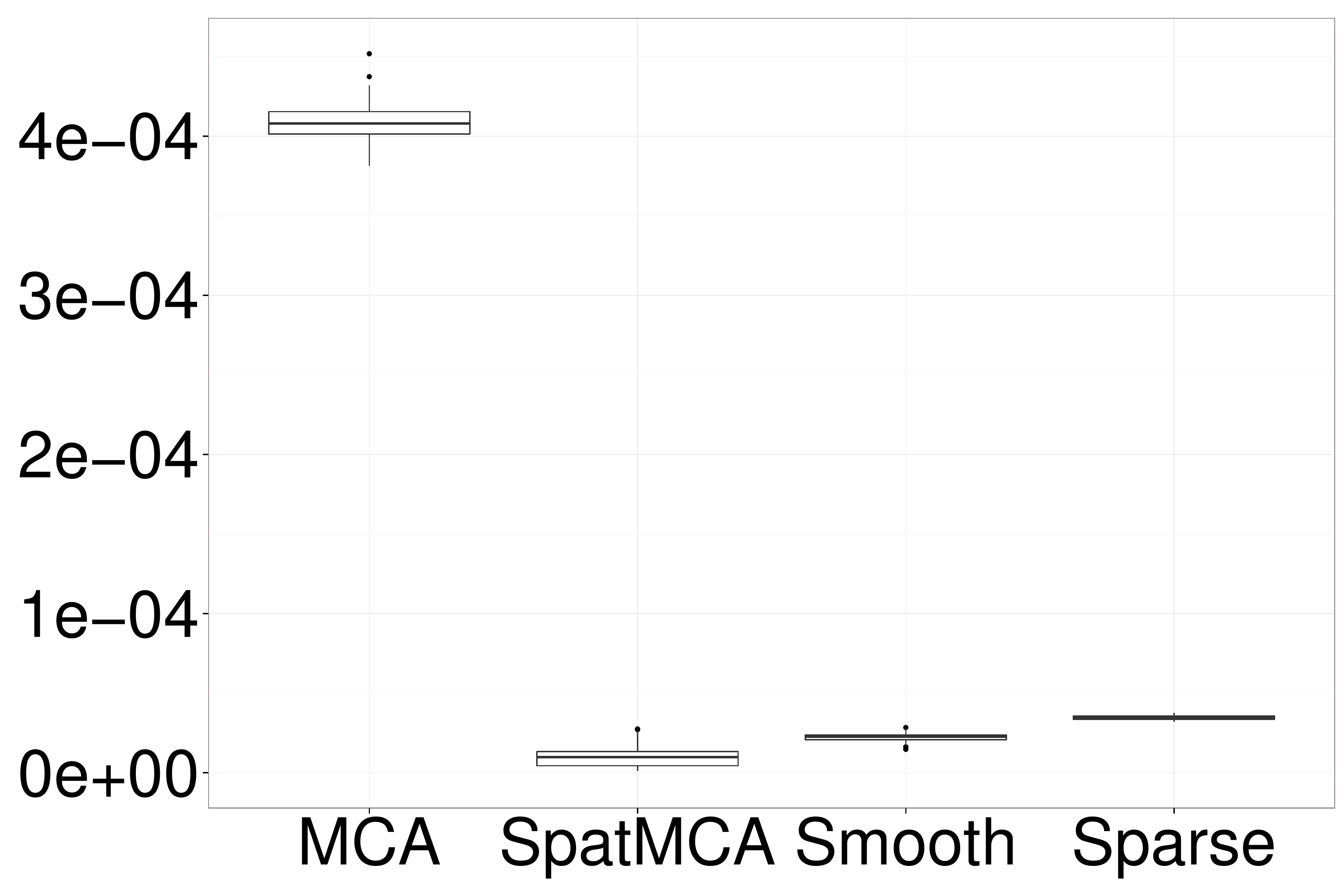}\hspace{4pt}&
		\includegraphics[scale=0.12]{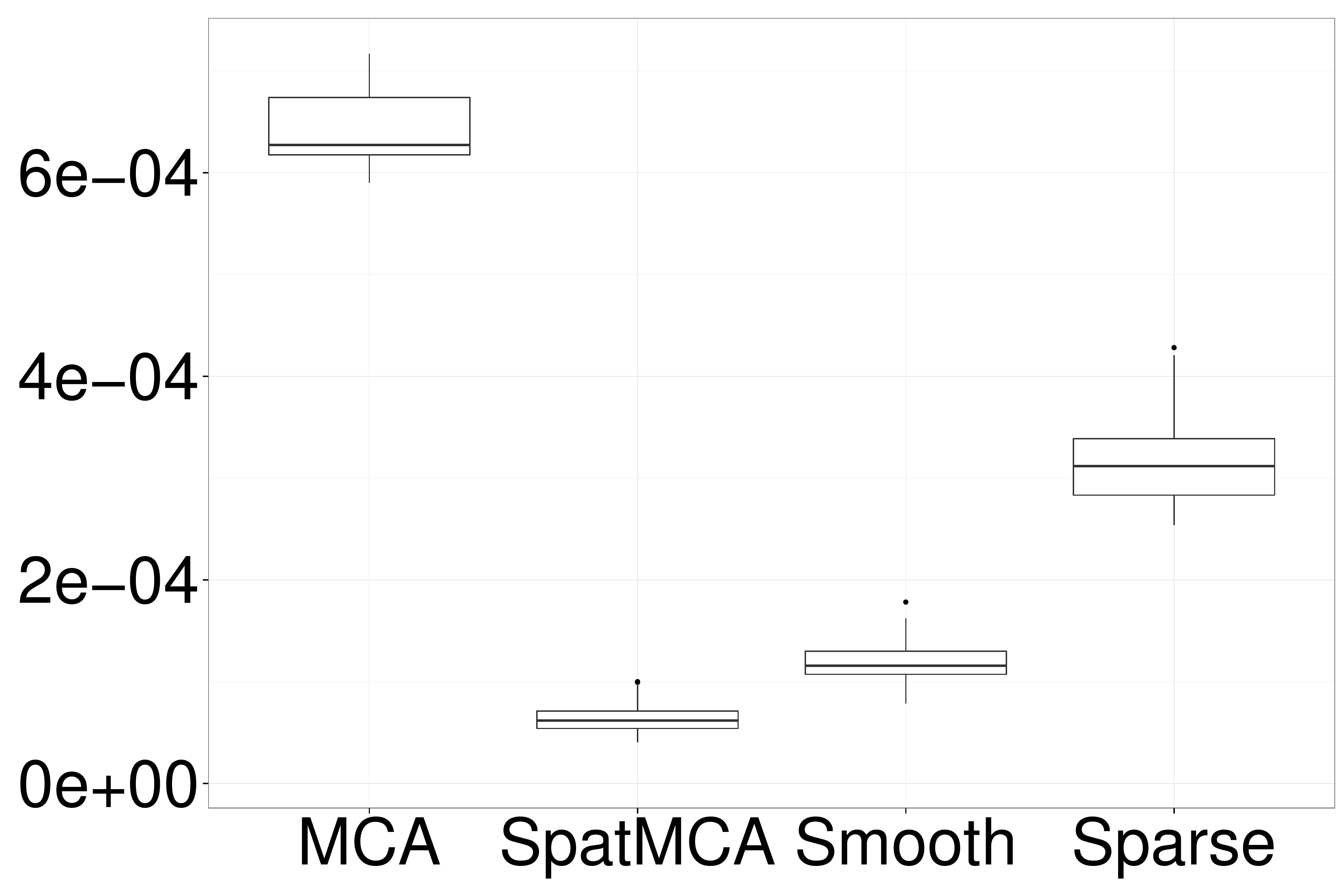}\hspace{4pt}\\
		{{$(d_1,d_2)=(1,0)$}, $K=\hat{K}$}&{{$(d_1,d_2)=(0.5,0)$}, $K=\hat{K}$}&{{$(d_1,d_2)=(1,0.7)$}, $K=\hat{K}$}\\
		\includegraphics[scale=0.12]{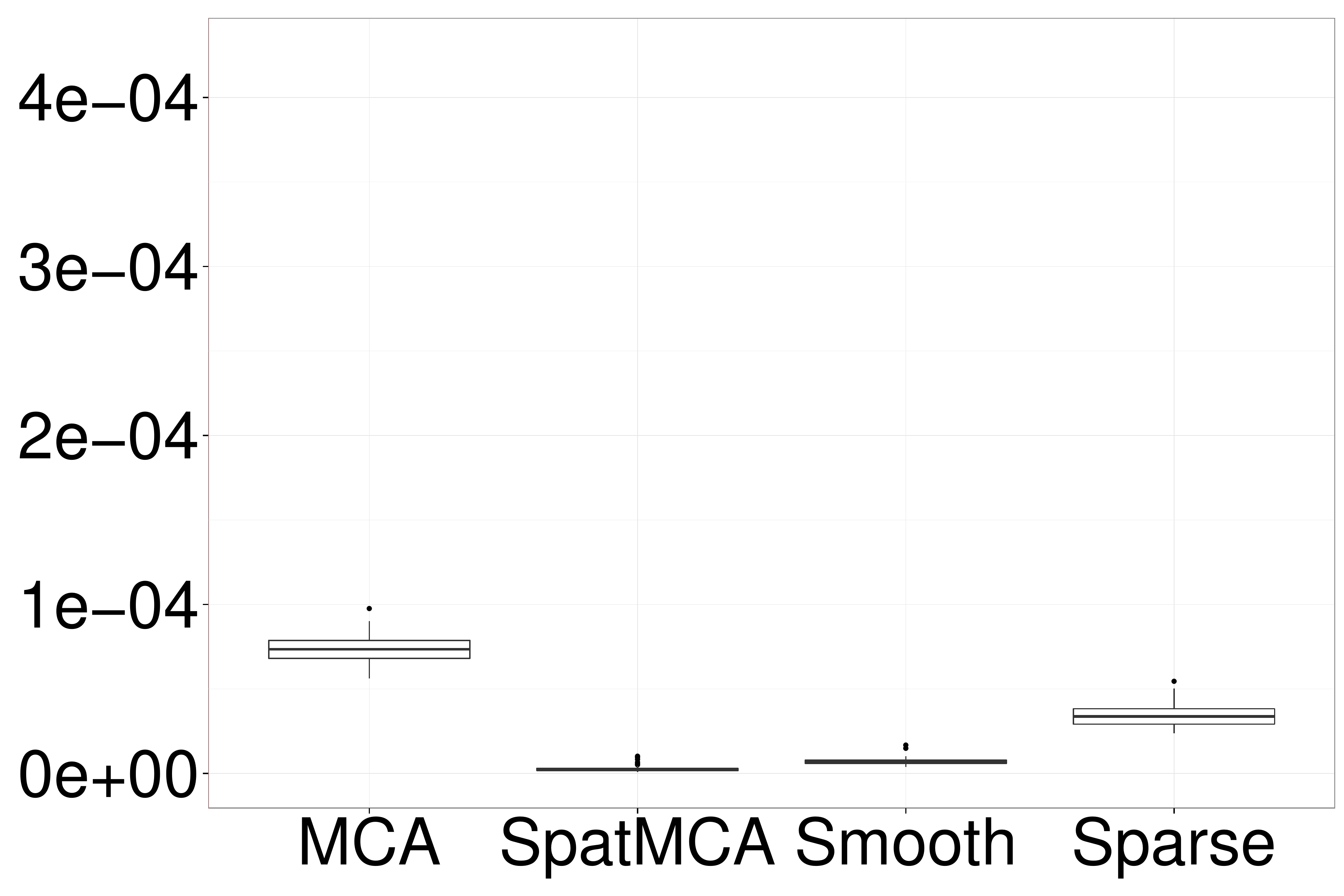}\hspace{4pt}&
		\includegraphics[scale=0.12]{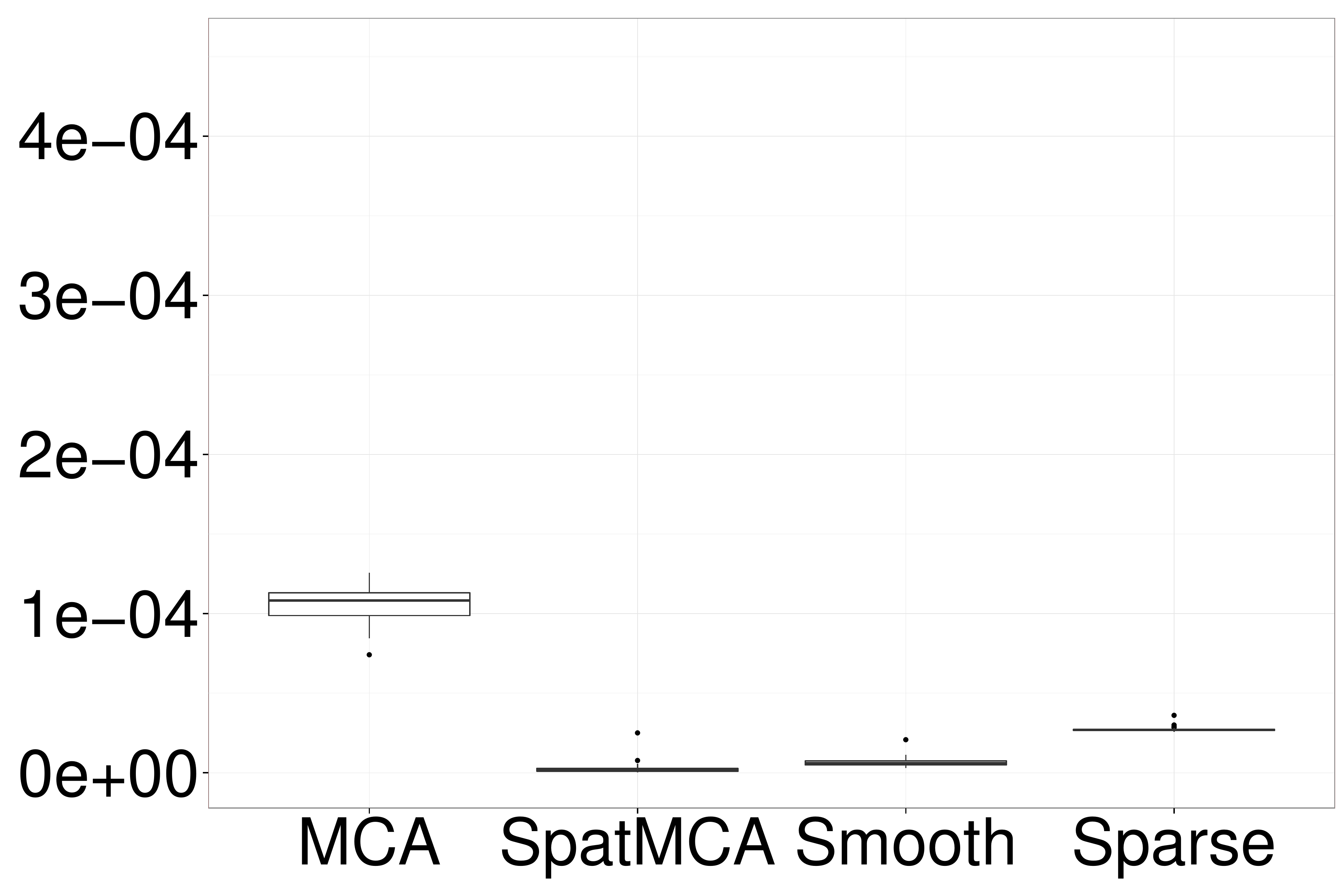}\hspace{4pt}&
		\includegraphics[scale=0.12]{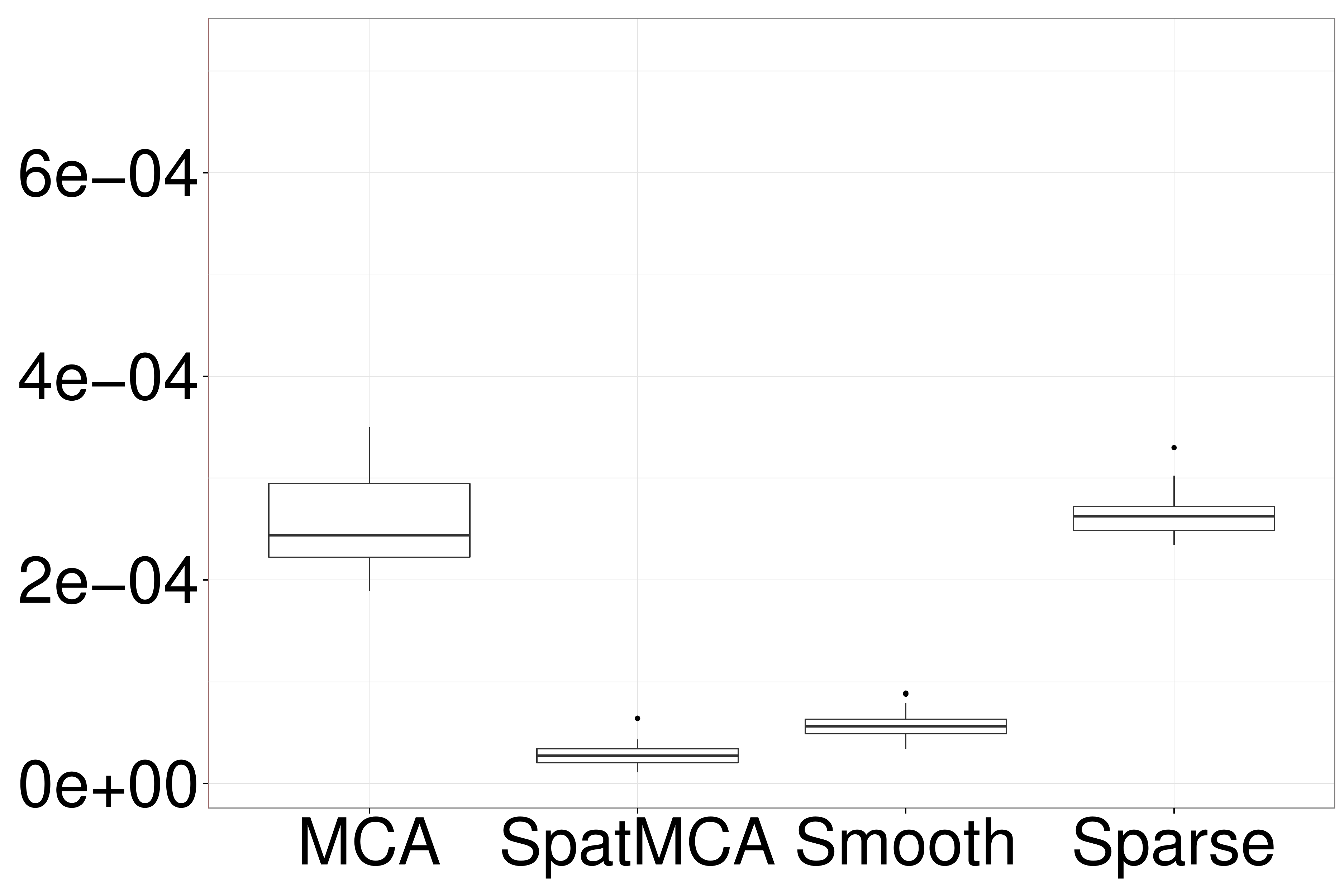}\hspace{4pt}
	\end{tabular}
	\caption{Boxplots of average squared prediction errors of (\ref{eq:loss1}) for various methods in the one-dimensional simulation experiment based on 50 simulation replicates.}
	\label{fig:box_d1_loss}
\end{figure}

The cross-covariance function estimates for the four methods based on a randomly generated dataset are shown in Figure~\ref{fig:cov_d1}. The proposed SpatMCA can be seen to perform better than the other methods for all cases. Figure~\ref{fig:box_d1_loss} shows boxplots of the four methods in terms of the loss function \eqref{eq:loss1} based on $50$ simulation replicates, which further confirms the superiority of SpatMCA.

\subsection{A Two-Dimensional Experiment}\label{sec:2d}

For a two-dimensional experiment, we generated data according to (\ref{eq:measurement}) with $K=2$, $n=5,000$, $p_1=25^2$, $p_2=20^2$,  \[\left(\begin{array}{c}
\bm{\eta}_{1i} \\
\bm{\eta}_{2i} \\
\end{array}\right)\sim N\left(\bm{0}, \left(\begin{array}{cc}
\bm{I} & \bm{U}\mathrm{diag}(d_1,d_2)\bm{V}' \\
\bm{V}\mathrm{diag}(d_1,d_2)\bm{U}' & \bm{I} \\
\end{array}\right)\right),\] 
$\bm{\epsilon}_{ji}\sim N(\bm{0},\bm{I})$ for $j=1,2$, $(\bm{s}_{11}, \dots,\bm{s}_{1p_1})$ equally spaced in $[-5,5]^2$, and $(\bm{s}_{21}, \dots,\bm{s}_{2p_2})$ equally spaced in $[-7,7]^2$.
Here $u_1(\cdot)$, $v_1(\cdot)$, $u_2(\cdot)$ {and} $v_2(\cdot)$ are given by \eqref{eq:u1_sim}, \eqref{eq:v1_sim}, \eqref{eq:u2_sim} and \eqref{eq:v2_sim} with $d=2$, respectively. We considered three pairs of $(d_1,d_2)\in \{(1,0), (0.5,0), (1,0.7)\}$, and applied the proposed SpatMCA with $K=\{1,2,5\}$ and $\hat{K}$ selected by \eqref{eq:ch5khat}, resulting in $12$ different combinations. Similar to the previous subsection, we applied {the} 5-fold CV of \eqref{eq:ch5khat} to select ${\{}\tau_{1u}, \tau_{1v}, \tau_{2u}, \tau_{2v}{\}}$ among $21$ values of $\tau_{1u}$ and $\tau_{1v}$  (including $0$ and the other 20 values equally spaced on the log scale from $10^{-2}$ to $10$) and $11$ values of $\tau_{2u}$ and $\tau_{2v}$ (including $0$ and the other 10 values equally spaced on the log scale from $10^{-3}$ to $1$). 

\begin{figure}\centering
	{$\hat{u}_1(\cdot)$ based on {$K=1$ for} $(d_1,d_2)=(1,0)$
		\includegraphics[scale=0.39]{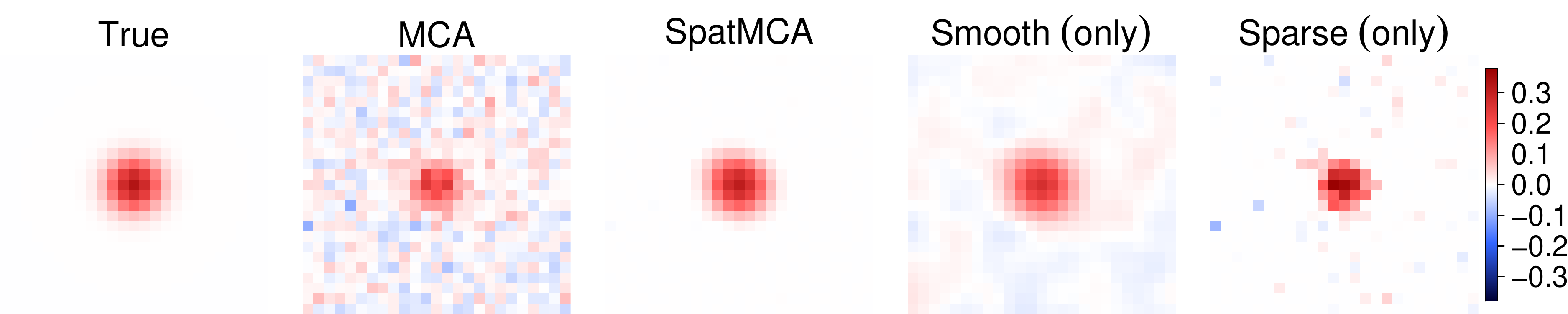}
		$\hat{u}_1(\cdot)$ based on {$K=1$ for} $(d_1,d_2)=(0.5,0)$
		\includegraphics[scale=0.39]{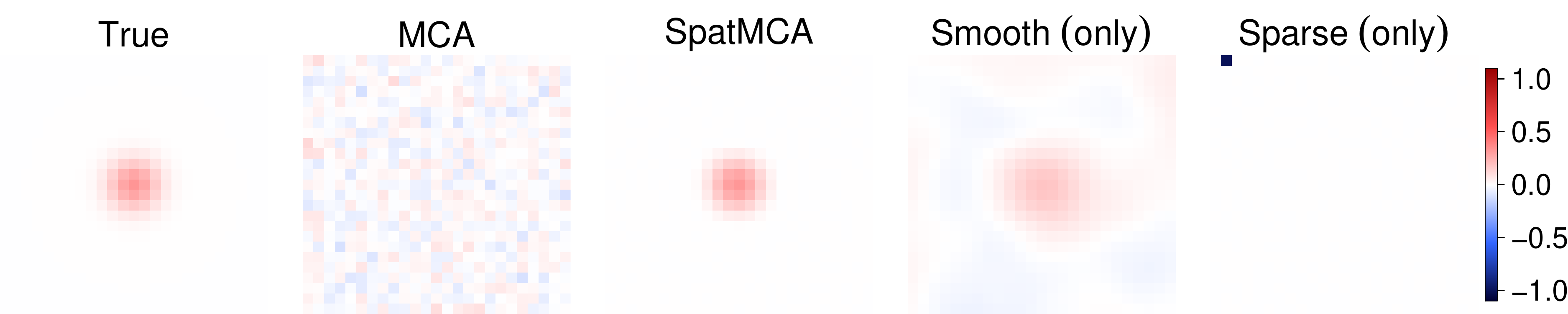}
		$\hat{u}_1(\cdot)$ based on {$K=2$ for} $(d_1,d_2)=(1,0.7)$
		\includegraphics[scale=0.39]{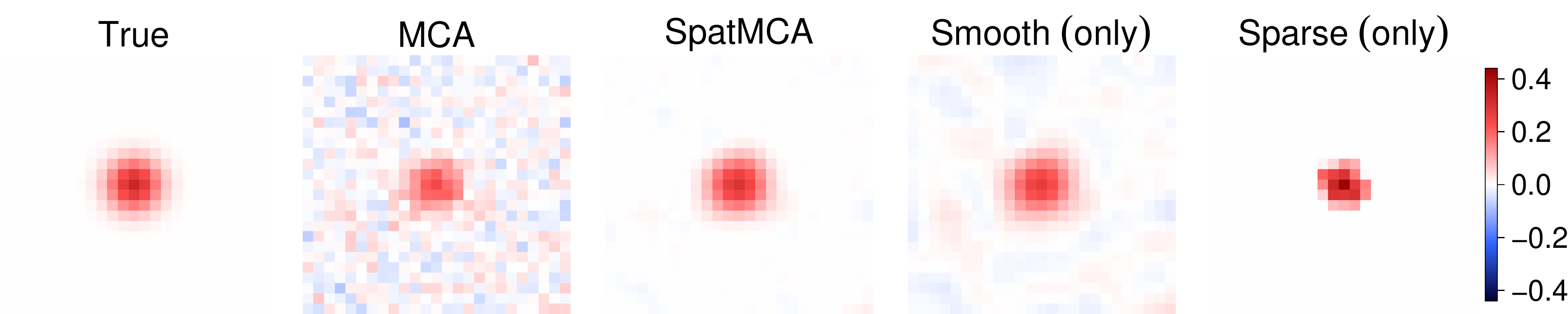}
		$\hat{u}_2(\cdot)$ based on {$K=2$ for} $(d_1,d_2)=(1,0.7)$}
	\includegraphics[scale=0.39]{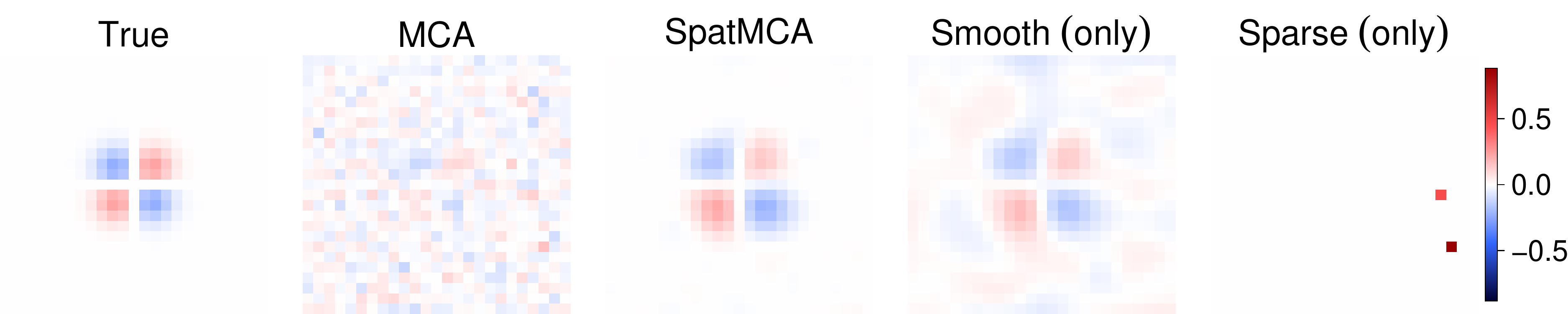}
	\caption{Estimates of $u_1(\cdot)$ and $u_2(\cdot)$ obtained from various methods based on three different combinations of singular values.}
	\label{fig:est_u_d2}
\end{figure}

\begin{figure}\centering
	$\hat{v}_1(\cdot)$ based on {$K=1$ for} $(d_1,d_2)=(1,0)$
	\includegraphics[scale=0.39]{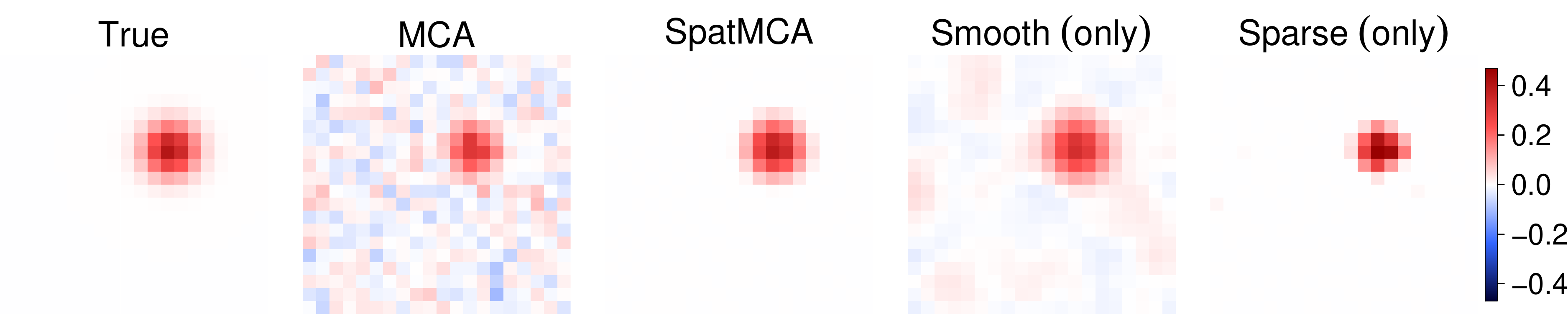}\\
	$\hat{v}_1(\cdot)$ based on {$K=1$ for} $(d_1,d_2)=(0.5,0)$
	\includegraphics[scale=0.39]{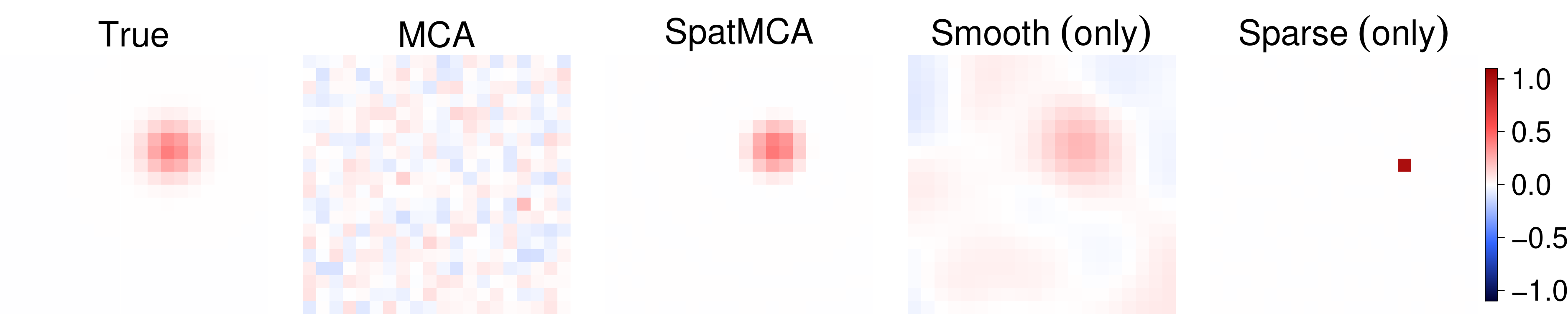}\\
	$\hat{v}_1(\cdot)$ based on {$K=2$ for} $(d_1,d_2)=(1,0.7)$
	\includegraphics[scale=0.39]{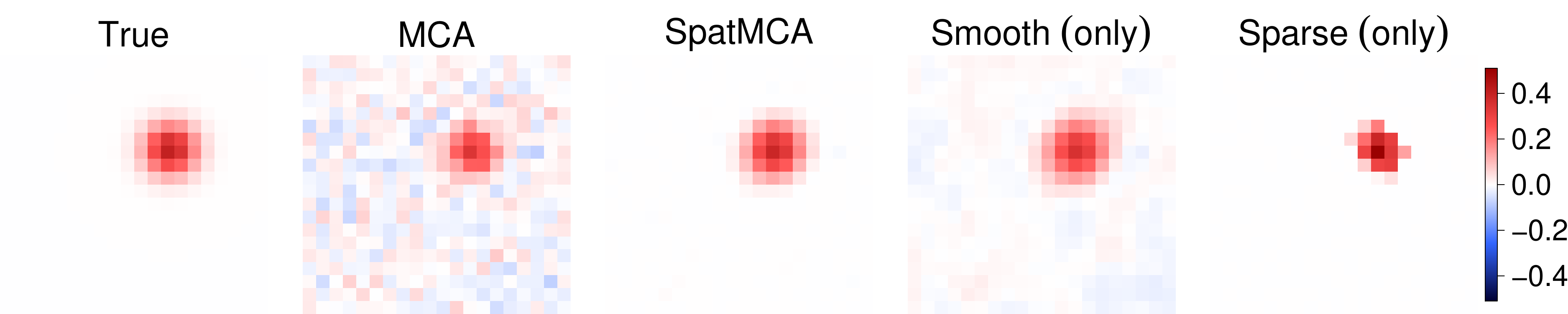}\\
	$\hat{v}_2(\cdot)$ based on {$K=2$ for} $(d_1,d_2)=(1,0.7)$
	\includegraphics[scale=0.39]{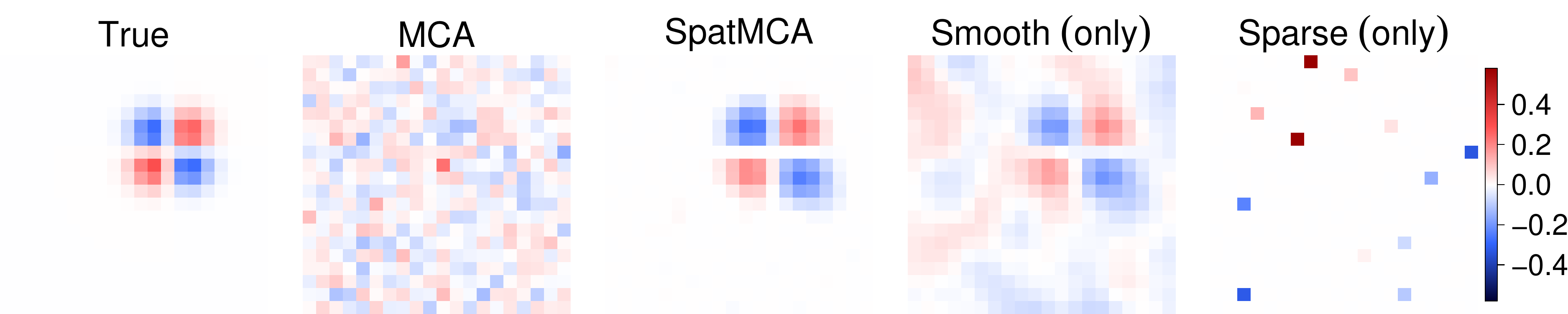}
	\caption{Estimates of $v_1(\cdot)$ and $v_2(\cdot)$ obtained from various methods based on three different combinations of singular values.}
	\label{fig:est_v_d2}
\end{figure}

\begin{figure}
	\begin{tabular}{ccc}
		{{$(d_1,d_2)=(1,0)$}, $K=1$}&{{$(d_1,d_2)=(0.5,0)$}, $K=1$}&{{$(d_1,d_2)=(1,0.7)$}, $K=1$}\\
		\includegraphics[scale=0.12]{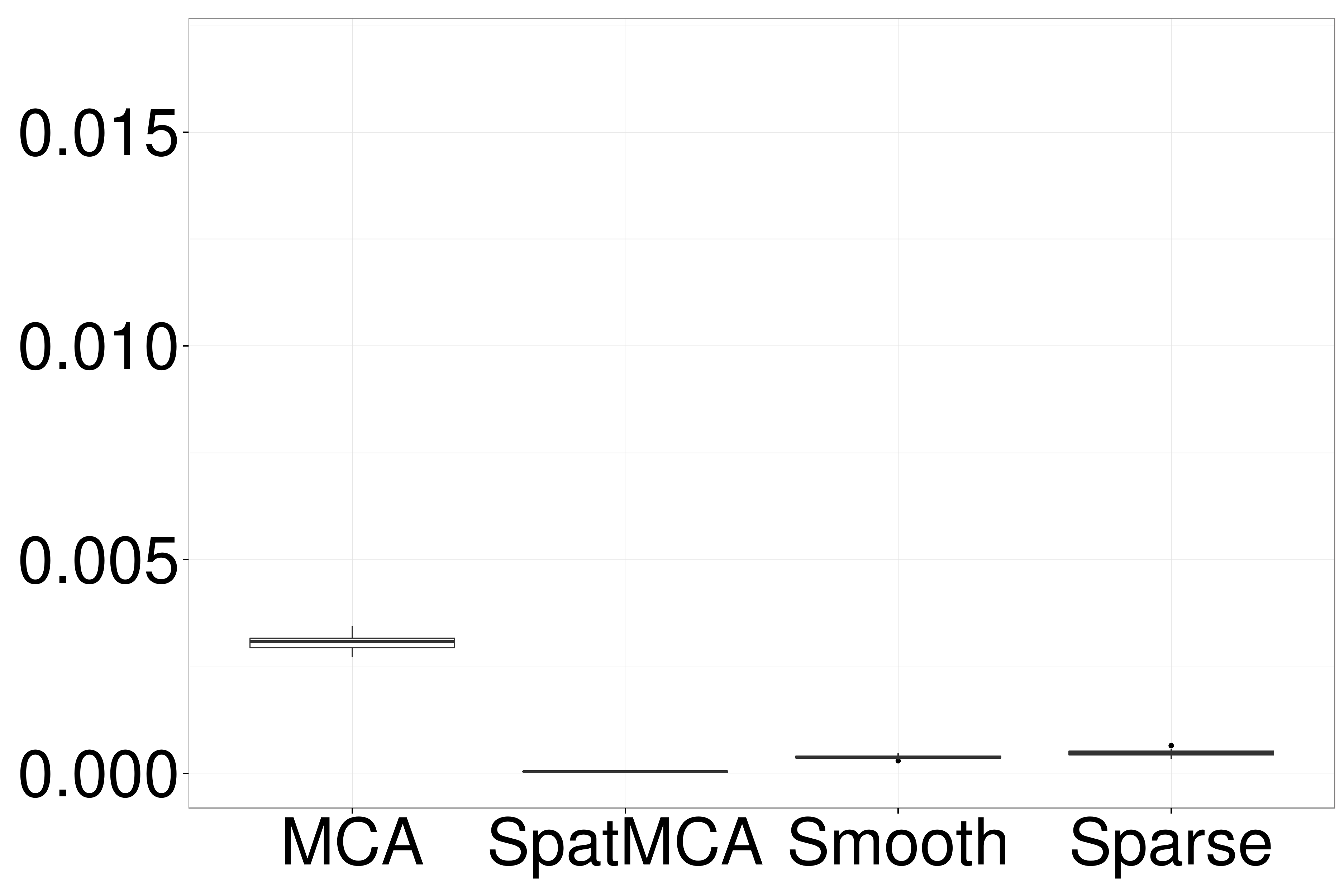}\hspace{4pt}&
		\includegraphics[scale=0.12]{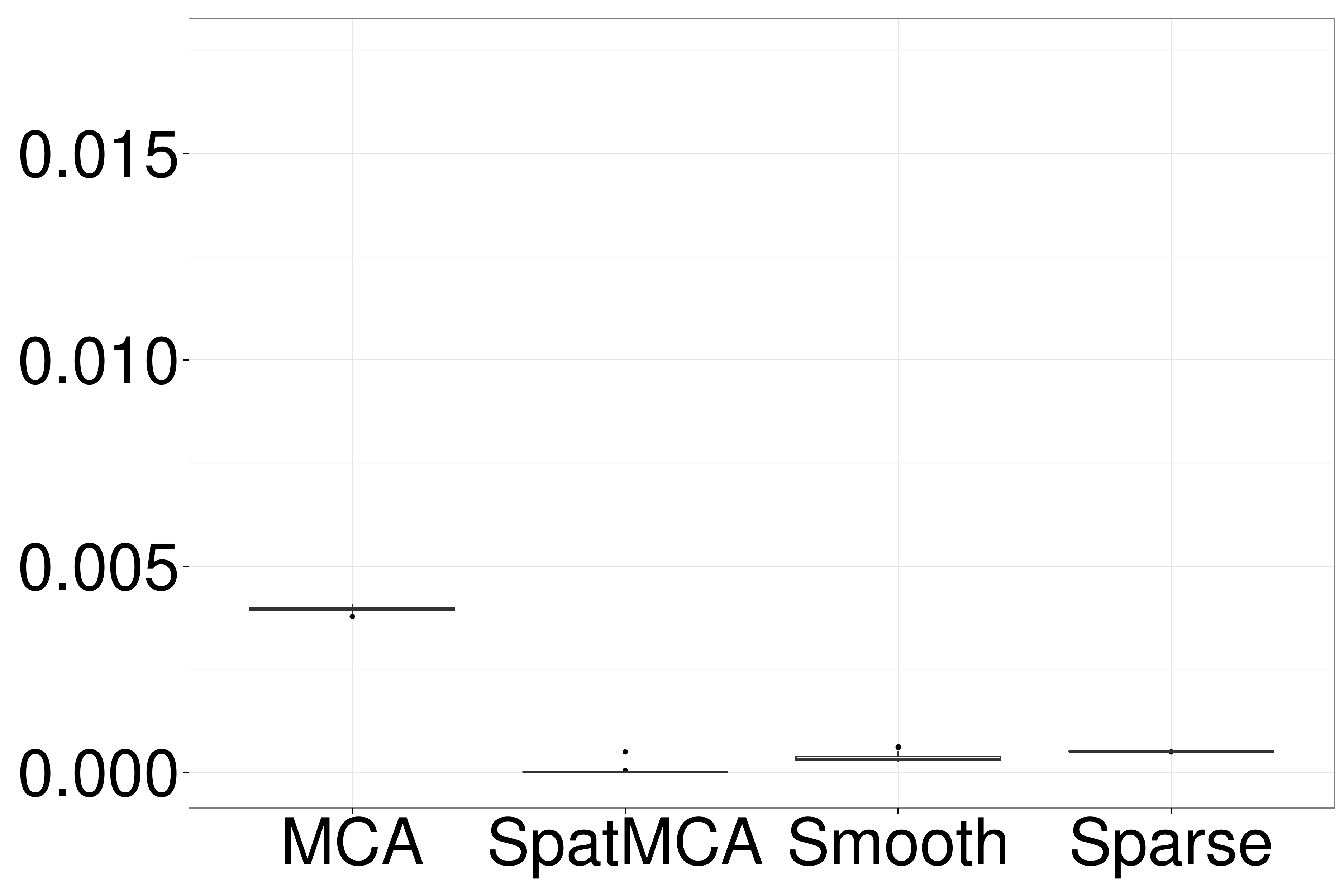}\hspace{4pt}&
		\includegraphics[scale=0.12]{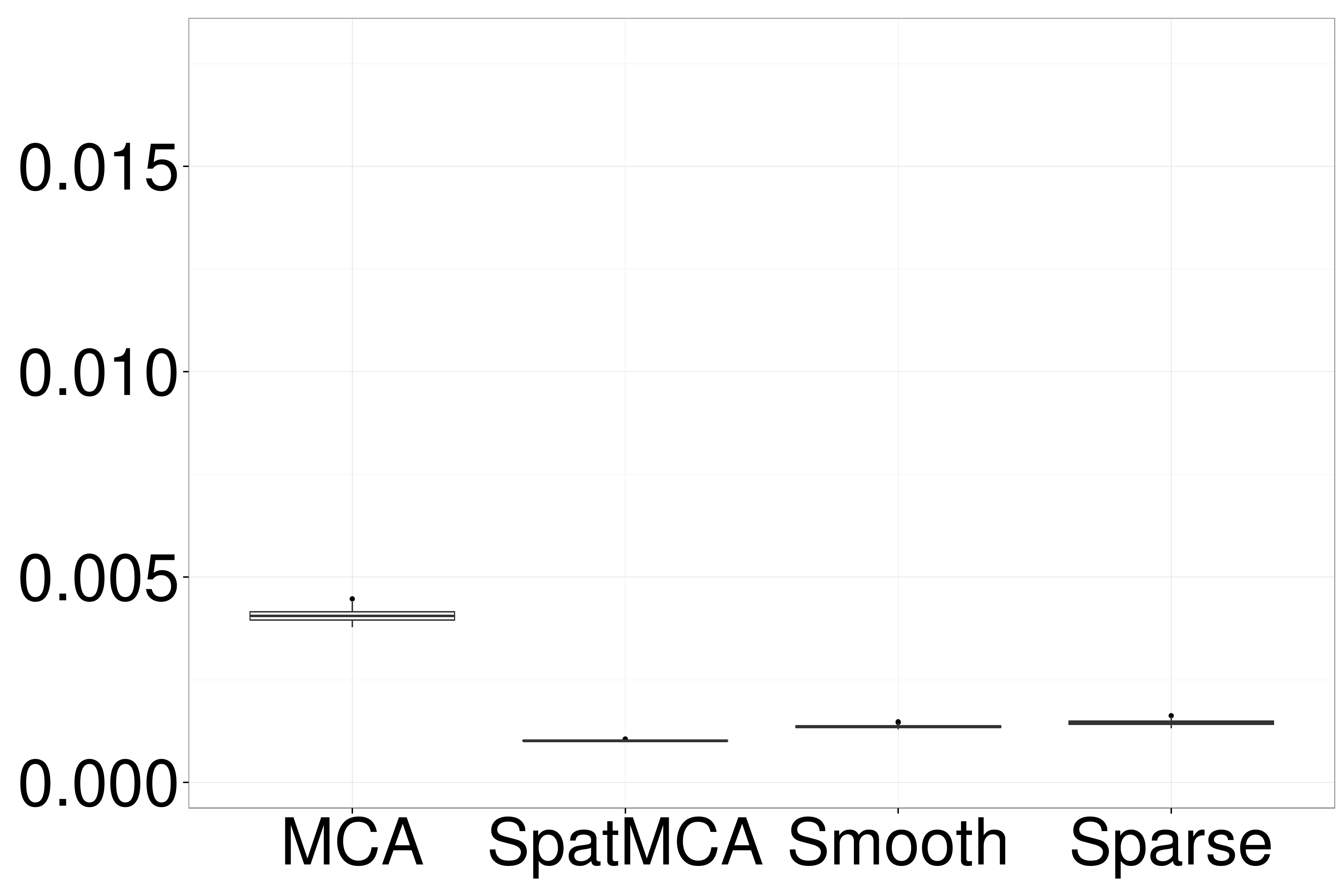}\hspace{4pt}\\
		{{$(d_1,d_2)=(1,0)$}, $K=2$}&{{$(d_1,d_2)=(0.5,0)$}, $K=2$}&{{$(d_1,d_2)=(1,0.7)$}, $K=2$}\\
		\includegraphics[scale=0.12]{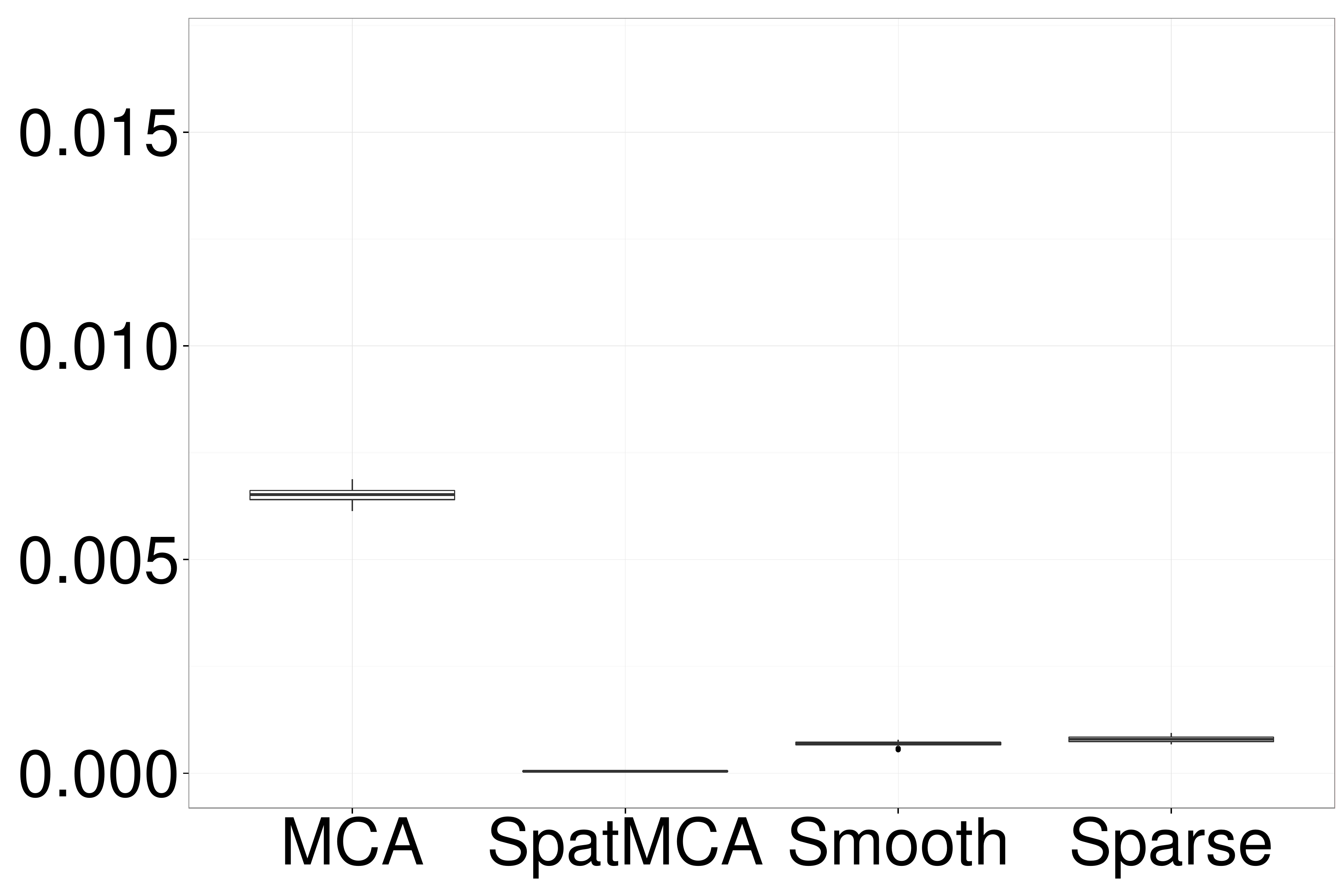}\hspace{4pt}&
		\includegraphics[scale=0.12]{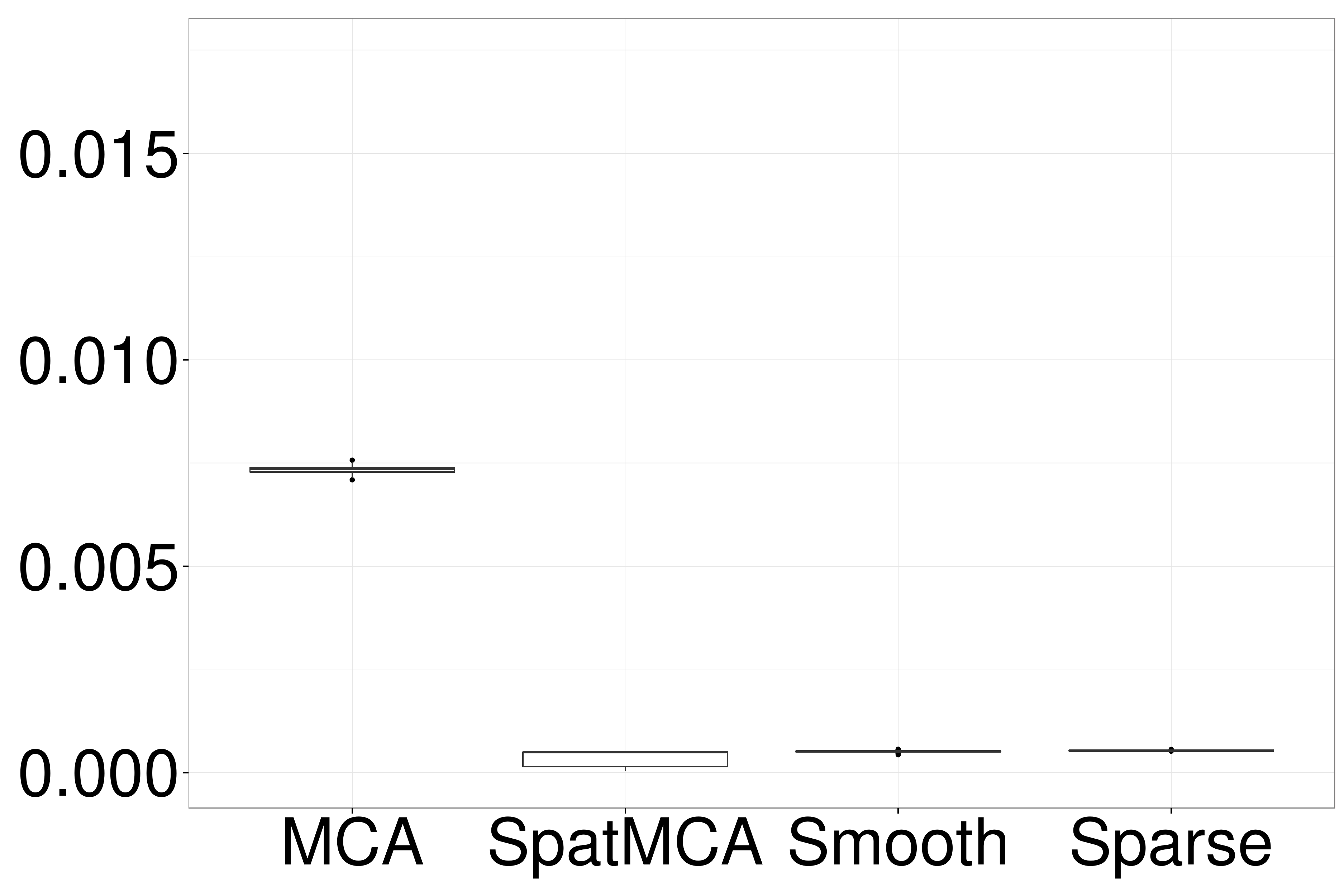}\hspace{4pt}&
		\includegraphics[scale=0.12]{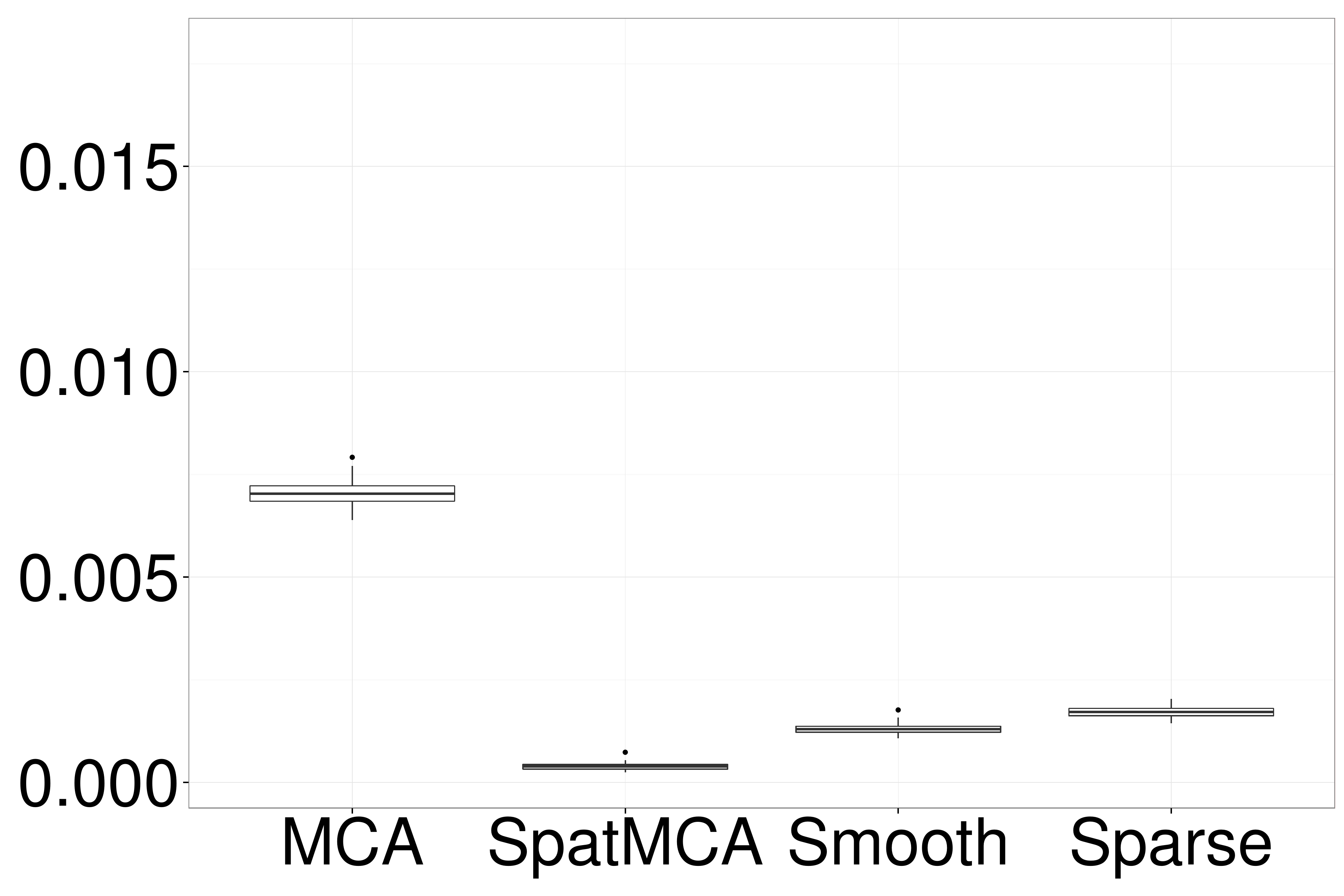}\hspace{4pt}\\
		{{$(d_1,d_2)=(1,0)$}, $K=5$}&{{$(d_1,d_2)=(0.5,0)$}, $K=5$}&{{$(d_1,d_2)=(1,0.7)$}, $K=5$}\\
		\includegraphics[scale=0.12]{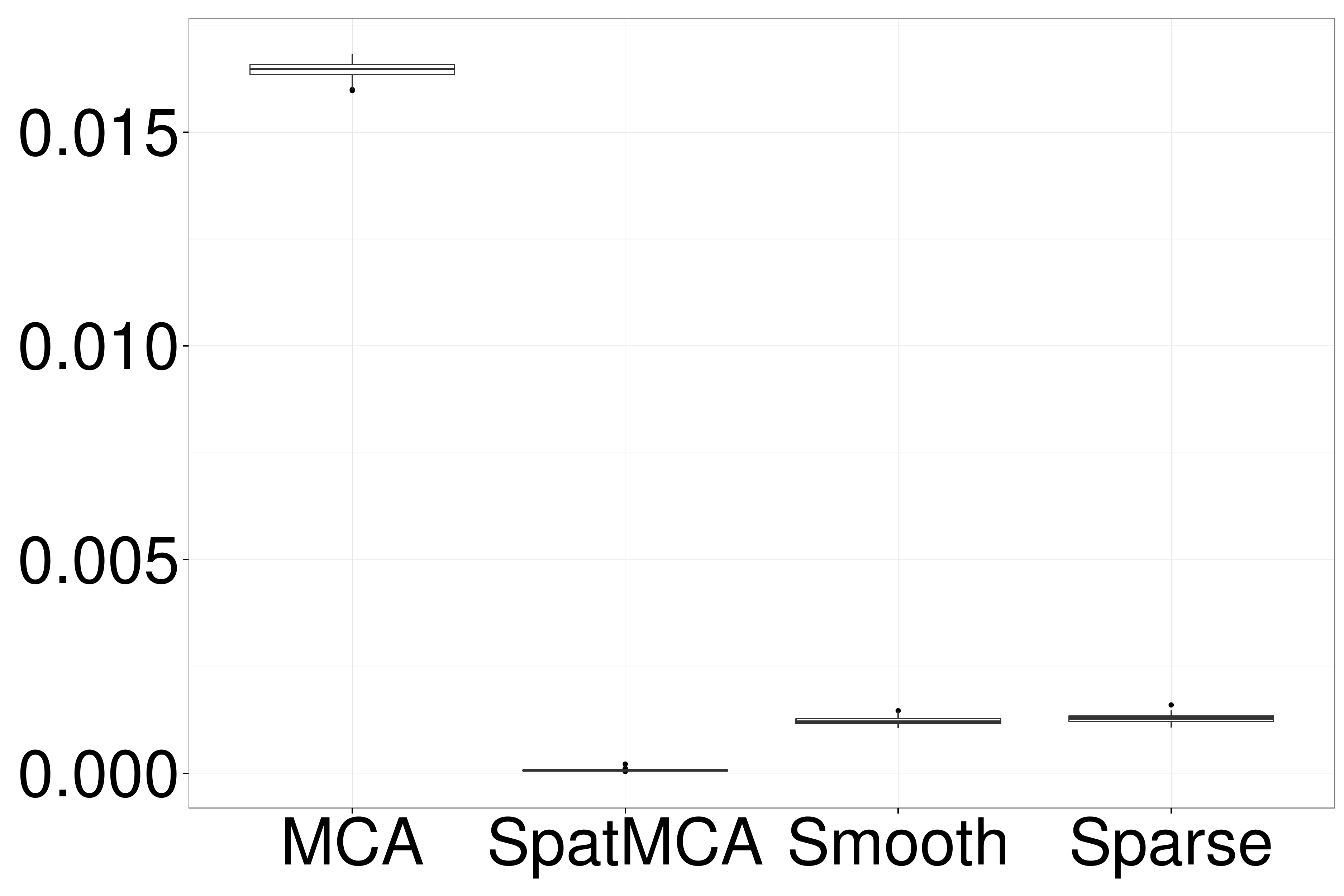}\hspace{4pt}&
		\includegraphics[scale=0.12]{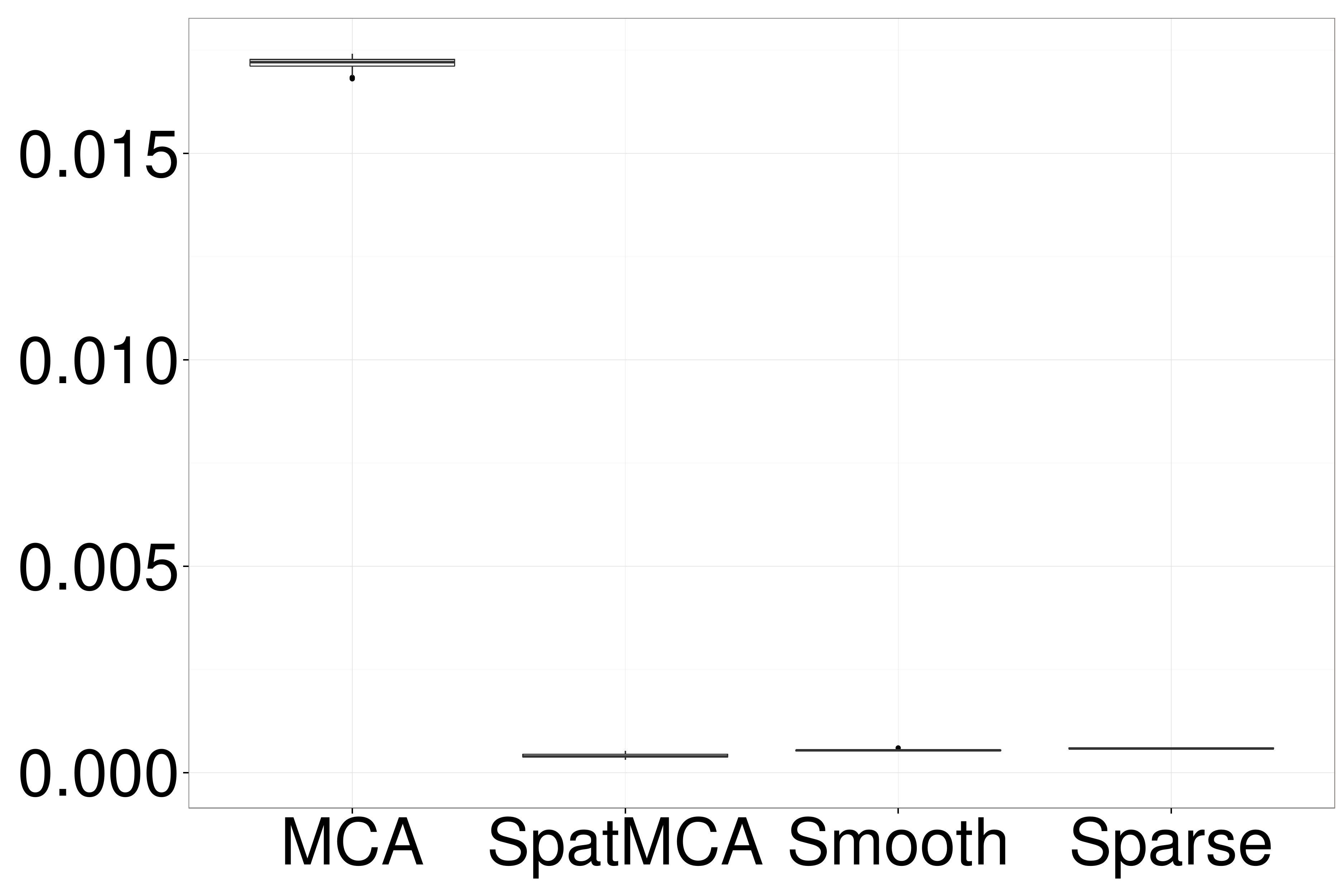}\hspace{4pt}&
		\includegraphics[scale=0.12]{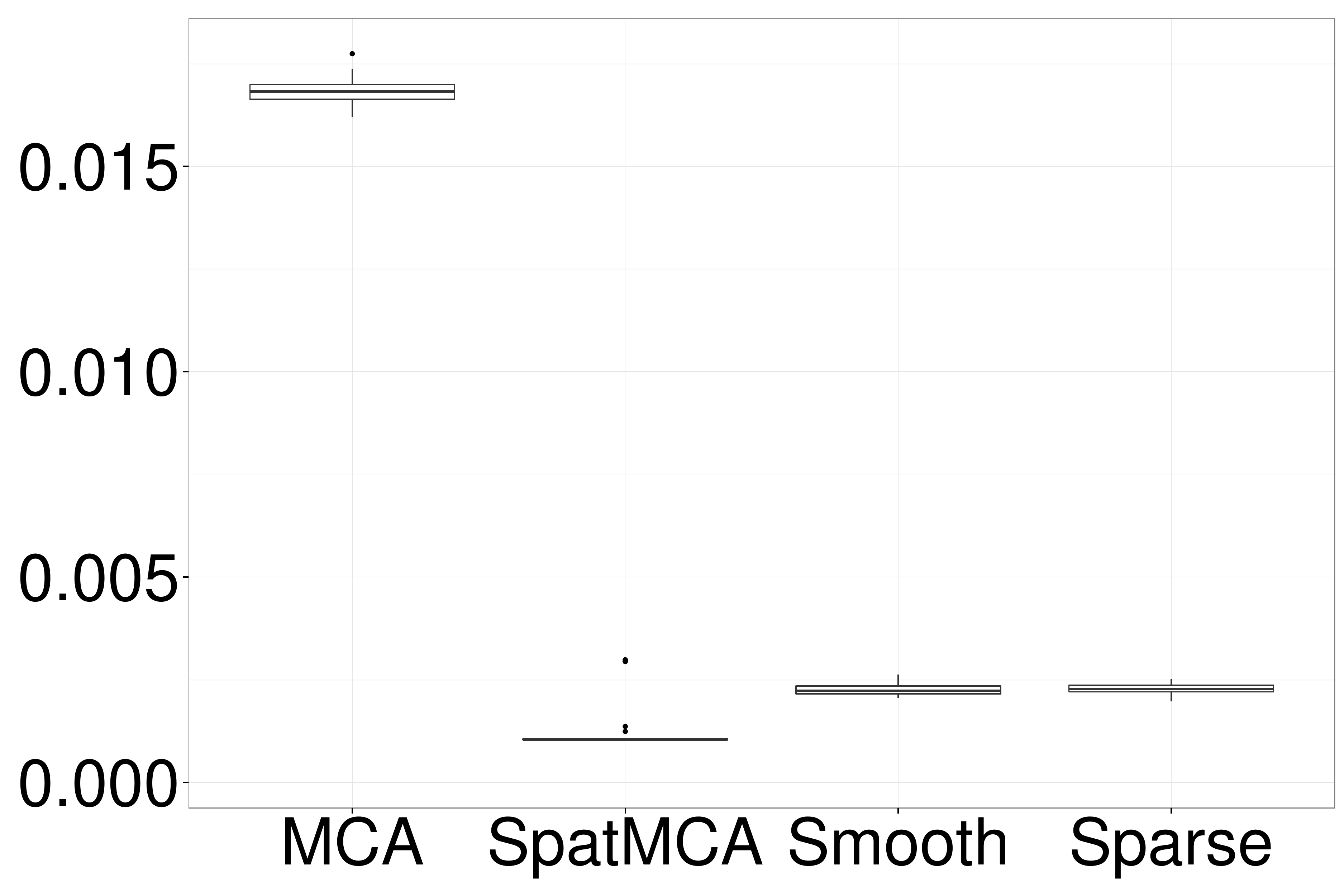}\hspace{4pt}\\
		{{$(d_1,d_2)=(1,0)$}, $K=\hat{K}$}&{{$(d_1,d_2)=(0.5,0)$}, $K=\hat{K}$}&{{$(d_1,d_2)=(1,0.7)$}, $K=\hat{K}$}\\
		\includegraphics[scale=0.12]{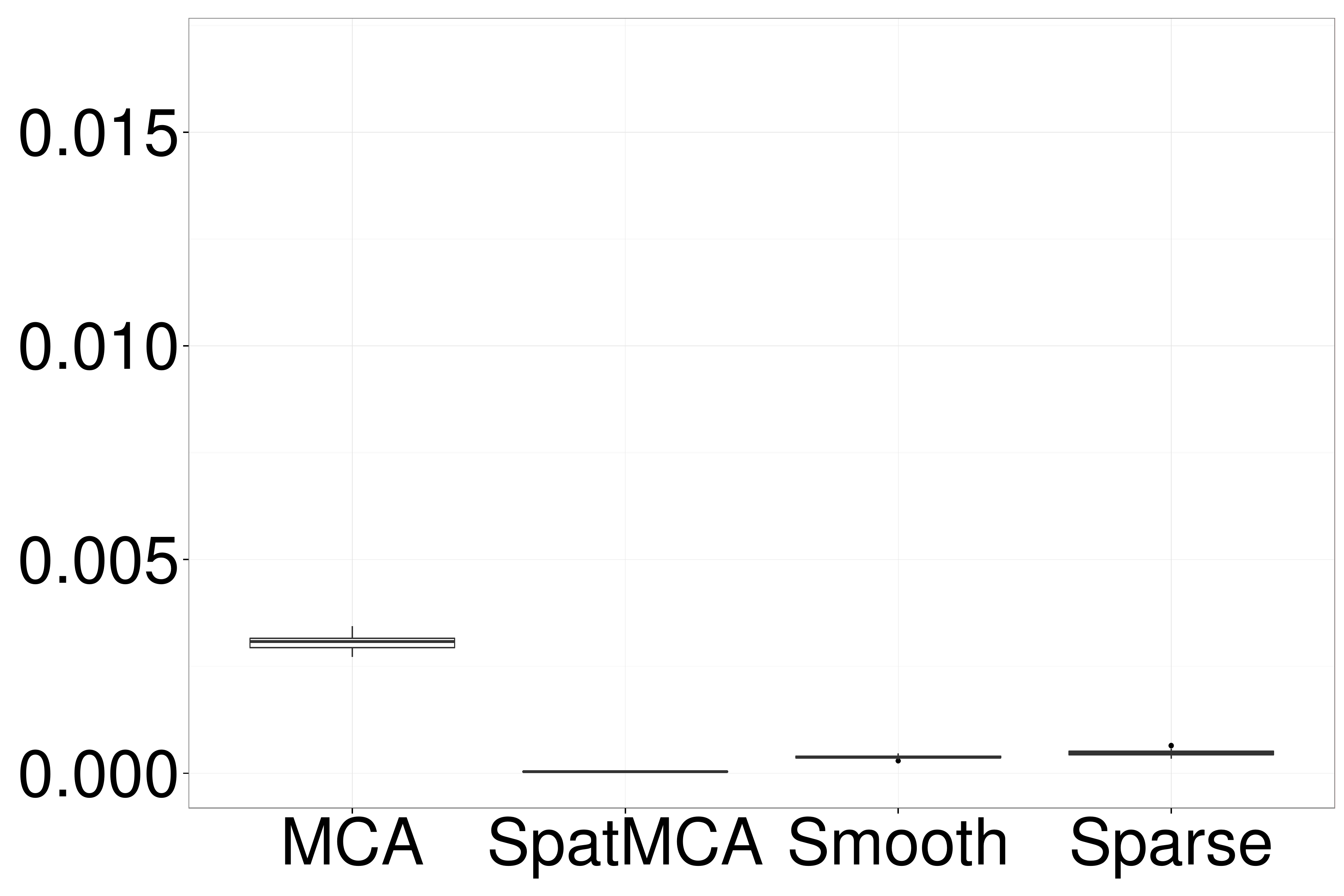}\hspace{4pt}&
		\includegraphics[scale=0.12]{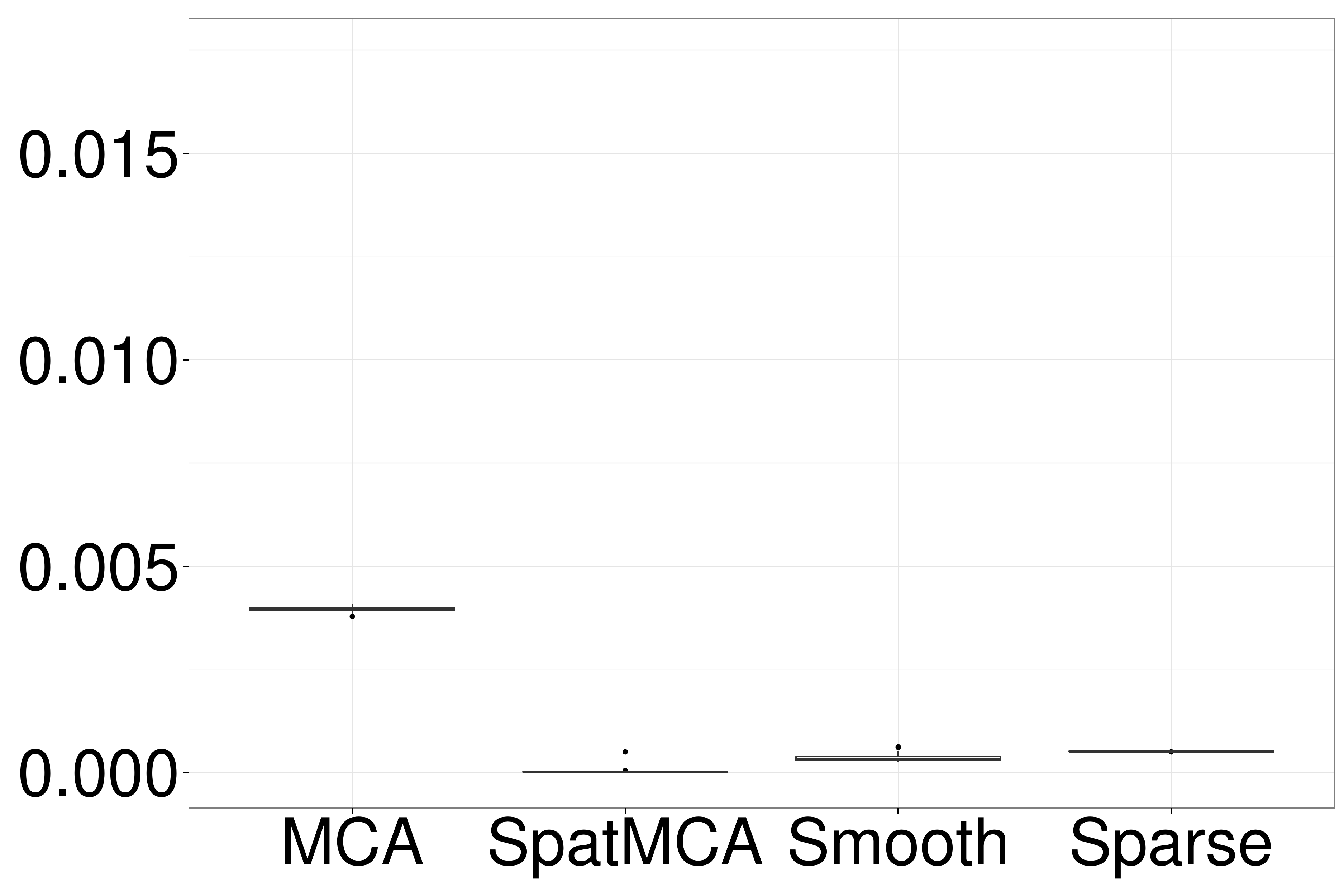}\hspace{4pt}&
		\includegraphics[scale=0.12]{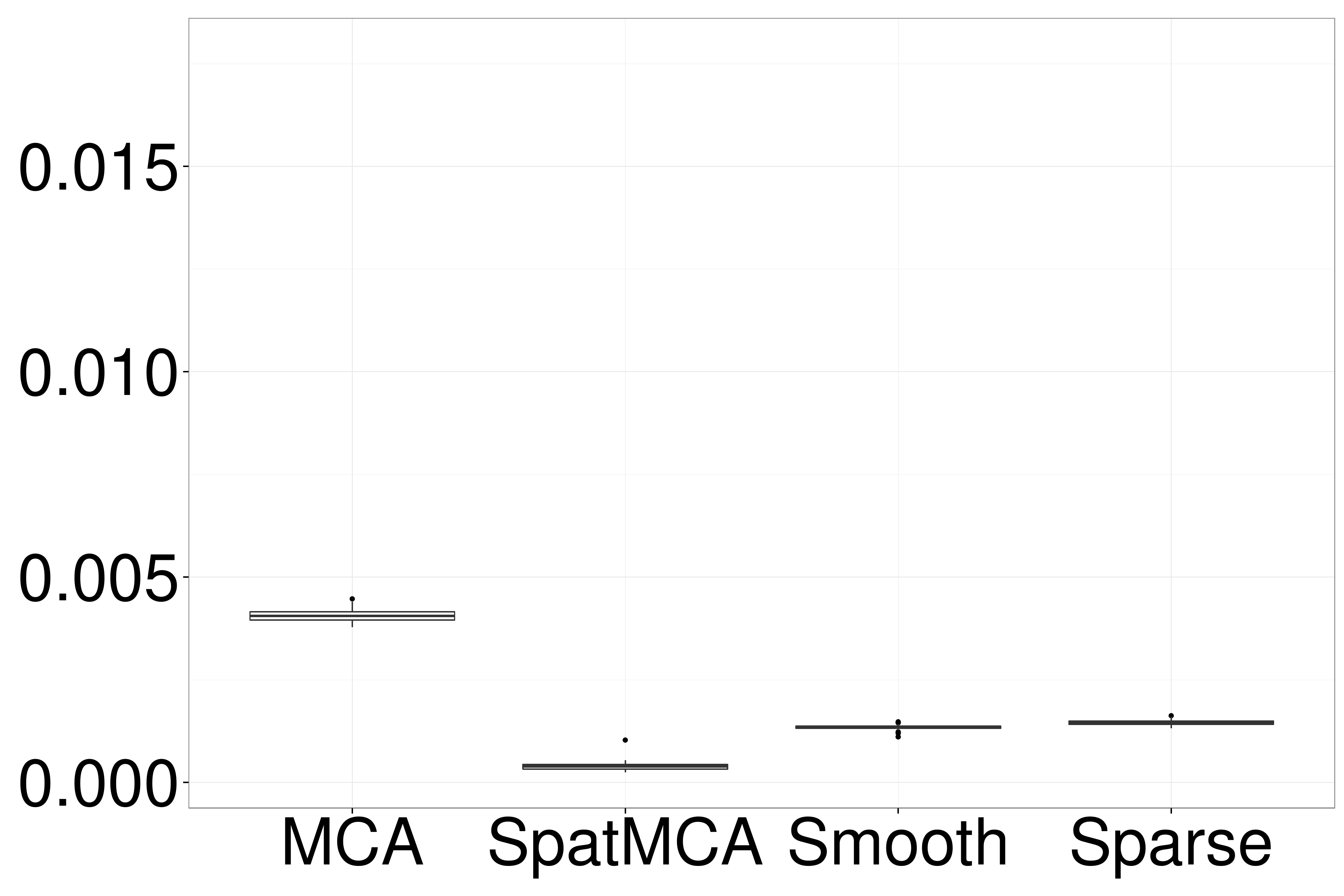}\hspace{4pt}
	\end{tabular}
	\caption{Boxplots of average squared prediction errors of (\ref{eq:loss1}) for various methods in the {two}-dimensional simulation experiment based on 50 simulation replicates.}
	\label{fig:box_d2_loss}
\end{figure}

Figures~\ref{fig:est_u_d2} and \ref{fig:est_v_d2} show the estimates of $u_k(\cdot)$ and $v_k(\cdot)$, respectively, for the four methods based on randomly selected data generated from three different combinations of singular values. Figure~\ref{fig:box_d2_loss} shows the performance of the four methods in terms of the loss function \eqref{eq:loss1} based on 50 simulation replicates. Similar to the one-dimensional example, SpatMCA outperforms all the other methods in all cases.

\subsection{An Application to Sea Surface Temperature and Precipitation Datasets}
We applied the proposed SpatMCA and MCA to investigate how precipitations in eastern Africa are affected by SSTs in the Indian Ocean and compared the differences between the two methods. The SST data are monthly averages (in degree Celsius) provided by the Met Office Marine Data Bank (available at \url{http://www.metoffice.gov.uk/hadobs/hadisst/}). The precipitation data are monthly averages (in  mm) provided by the Earth System Research Laboratory, Physical Science Division of the National Oceanic and Atmospheric Administration (available at \url{http://www.esrl.noaa.gov/psd/}). Both datasets are on $1$ degree latitude by 1 degree longitude equiangular grid cells. As in \citet{omondi2013influence}, we considered a region of the Indian Ocean between latitudes $20^{\circ}$N and $30^{\circ}$S and between longitudes $20^{\circ}$E and $120^{\circ}$E for the SST dataset, and we chose a region of eastern Africa between $6^{\circ}$N and $12^{\circ}$S and between longitudes $20^{\circ}$E and $42^{\circ}$E for the precipitation dataset. We used the data observed from January 2011 to December 2015. Let $\bm{\eta}_{1i}$ and $\bm{\eta}_{2i}$ be the vectors of \eqref{eq:measurement} corresponding to SST in the Indian Ocean and precipitation in eastern Africa. In this example, $p_1 = 3,591$, $p_2=255$, and $n=60$.

First, the SST data and the precipitation data were detrended by subtracting their individual average for a given cell and a given month. Then, the data were randomly split into two parts as the training data and  the validation data. We applied SpatMCA to the training data with $K$ selected by $\hat{K}$ of $(\ref{eq:ch5khat})$, where 21 values of $\tau_{1u}$ and $\tau_{1v}$ (including $0$ and the other 20 values equally spaced on the log scale from $10^{-1}$ to $10^6$) and 21 values of $\tau_{2u}$ and $\tau_{2v}$ (including $0$ and the other $20$ values equally spaced on the log scale from $10^{-3}$ to $0.5$) were selected by using 5-fold CV of \eqref{eq:cv_max1} and \eqref{eq:cv_max2}. 

\begin{figure}
	\centering
	\includegraphics[height=5cm, width=6 cm]{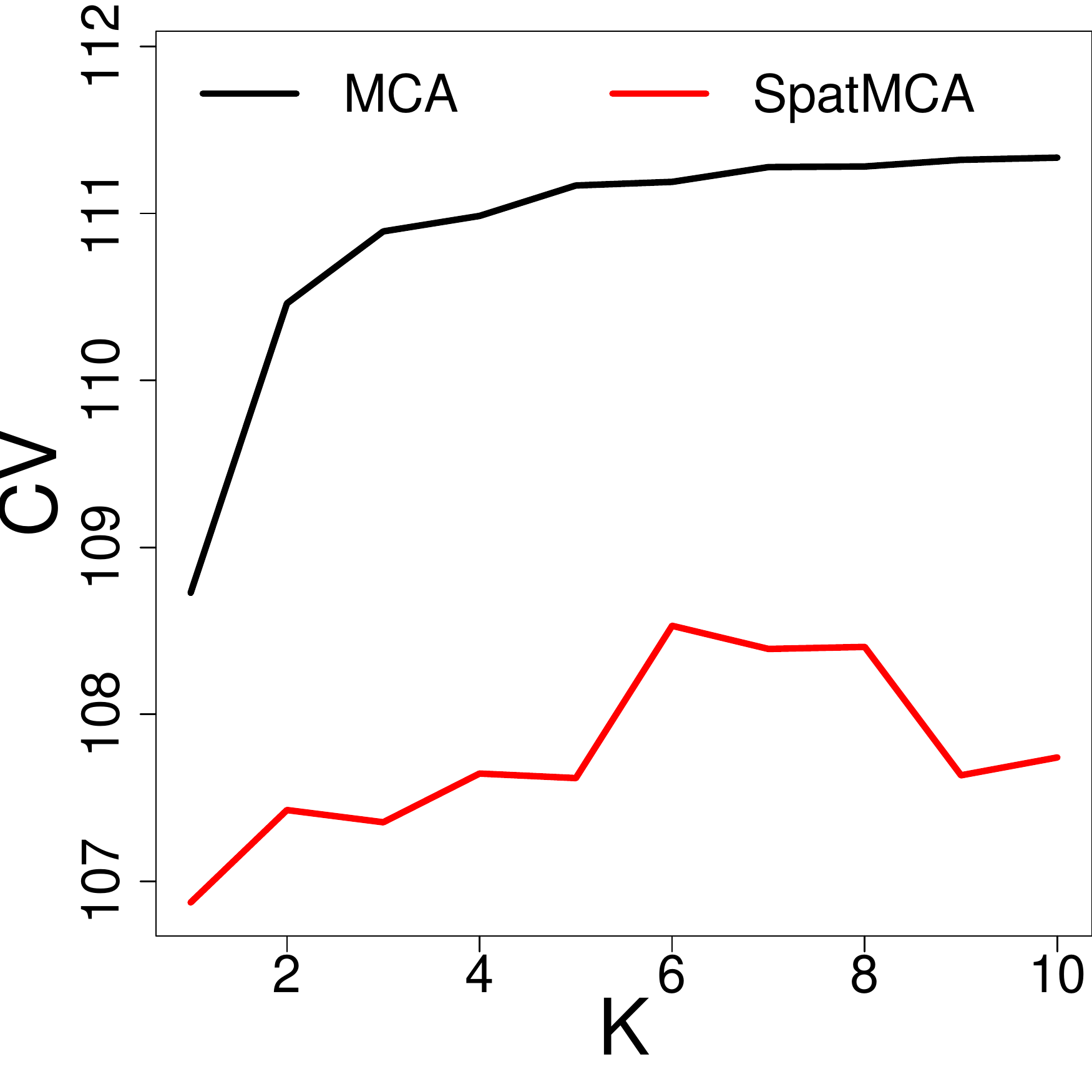}
	\caption{CV values with respect to $K$ for SpatMCA and MCA.}
	\label{fig:cv_real}
\end{figure}

\begin{figure}\centering
	$\hat{u}_1(\cdot)$ from MCA\hspace{120pt}$\hat{u}_1(\cdot)$ from SpatMCA
	\includegraphics[scale=0.2]{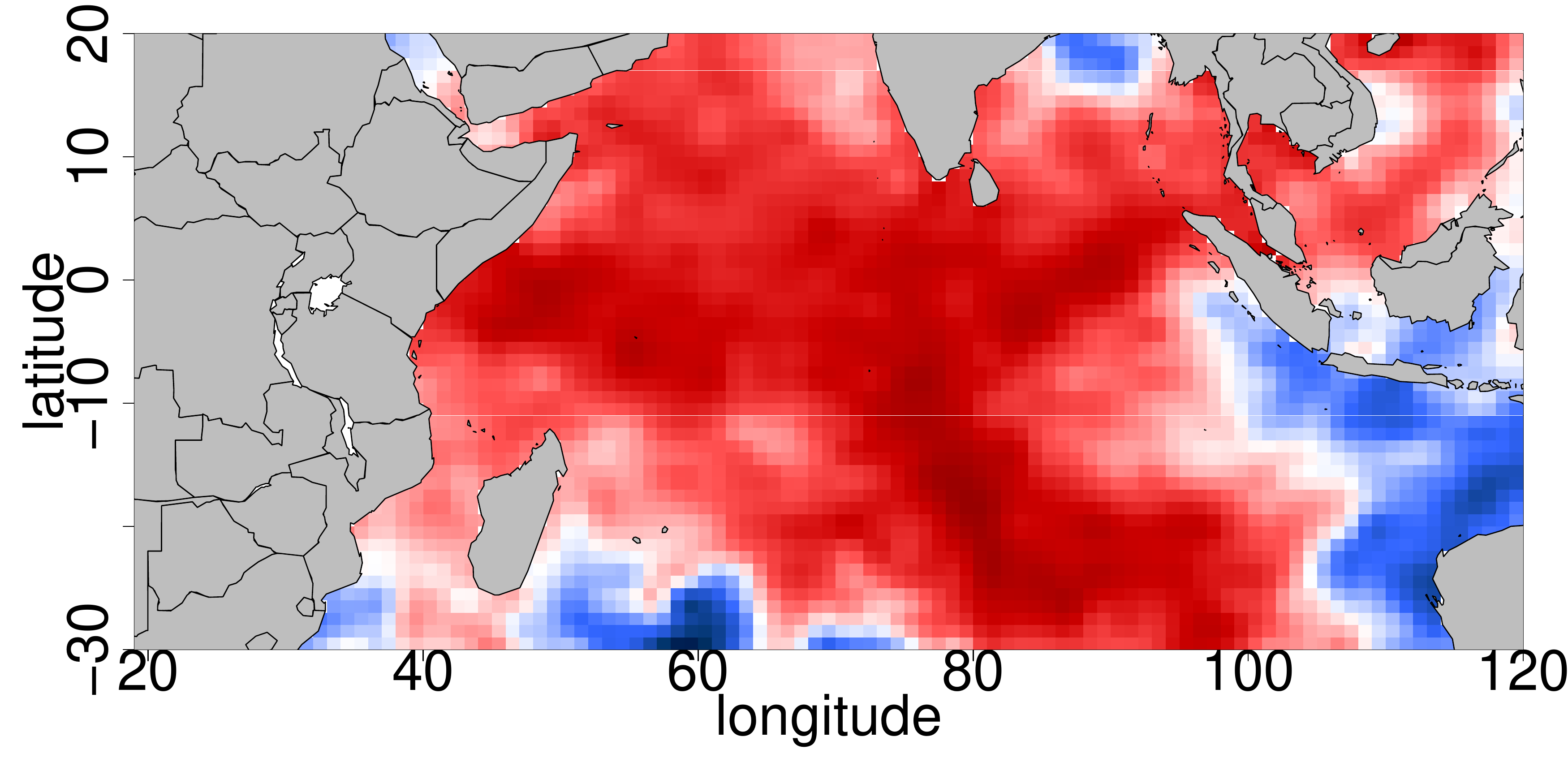}
	\includegraphics[scale=0.2]{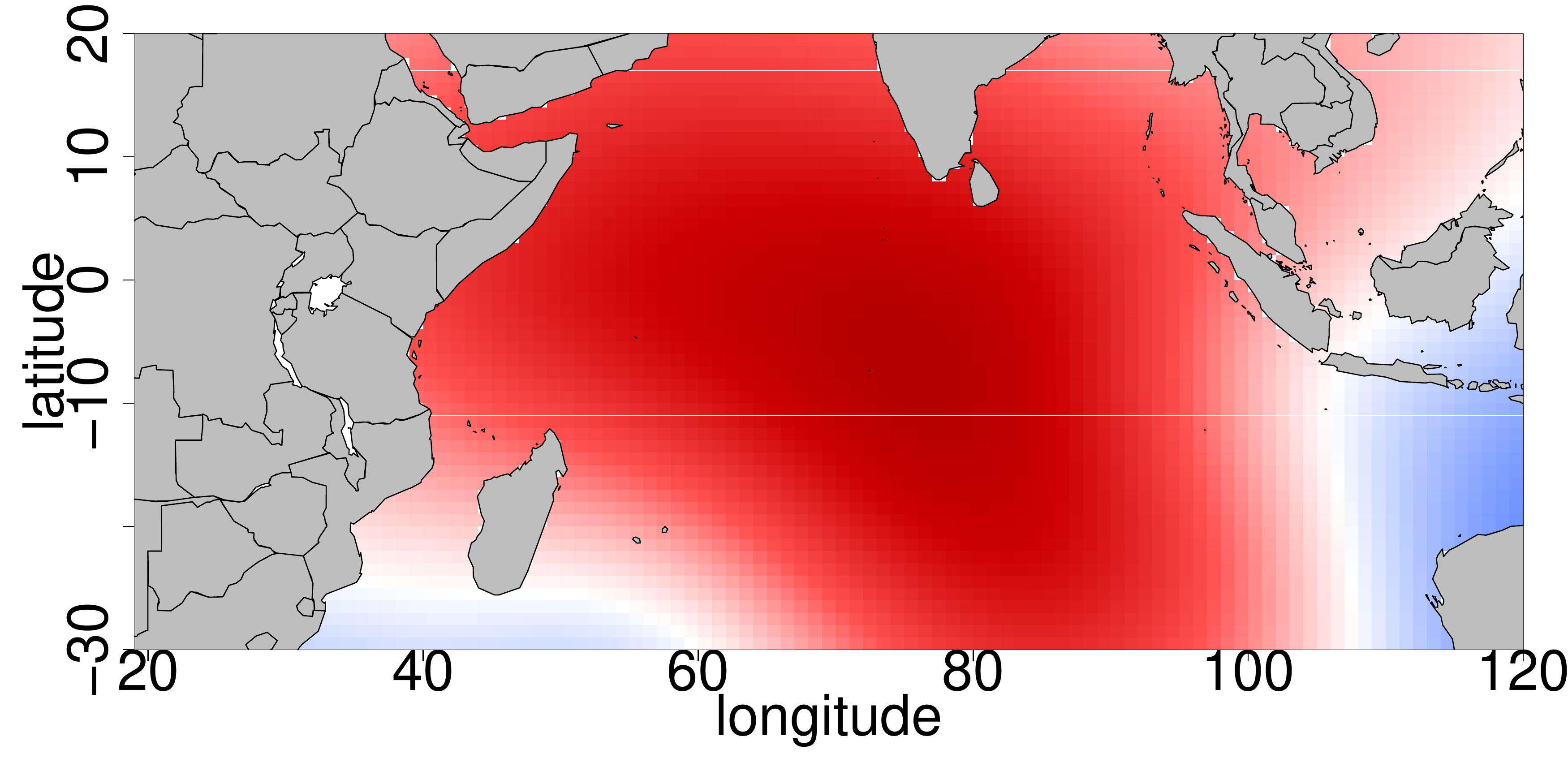}\\
	{\includegraphics[scale=0.2]{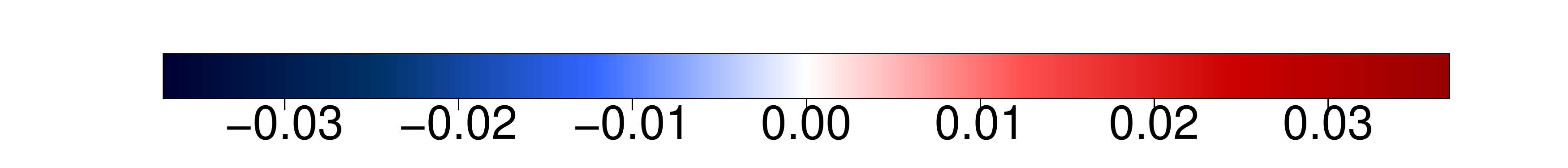}}\\
	$\hat{v}_1(\cdot)$ from MCA\hspace{120pt}$\hat{v}_1(\cdot)$ from SpatMCA
	\includegraphics[scale=0.2]{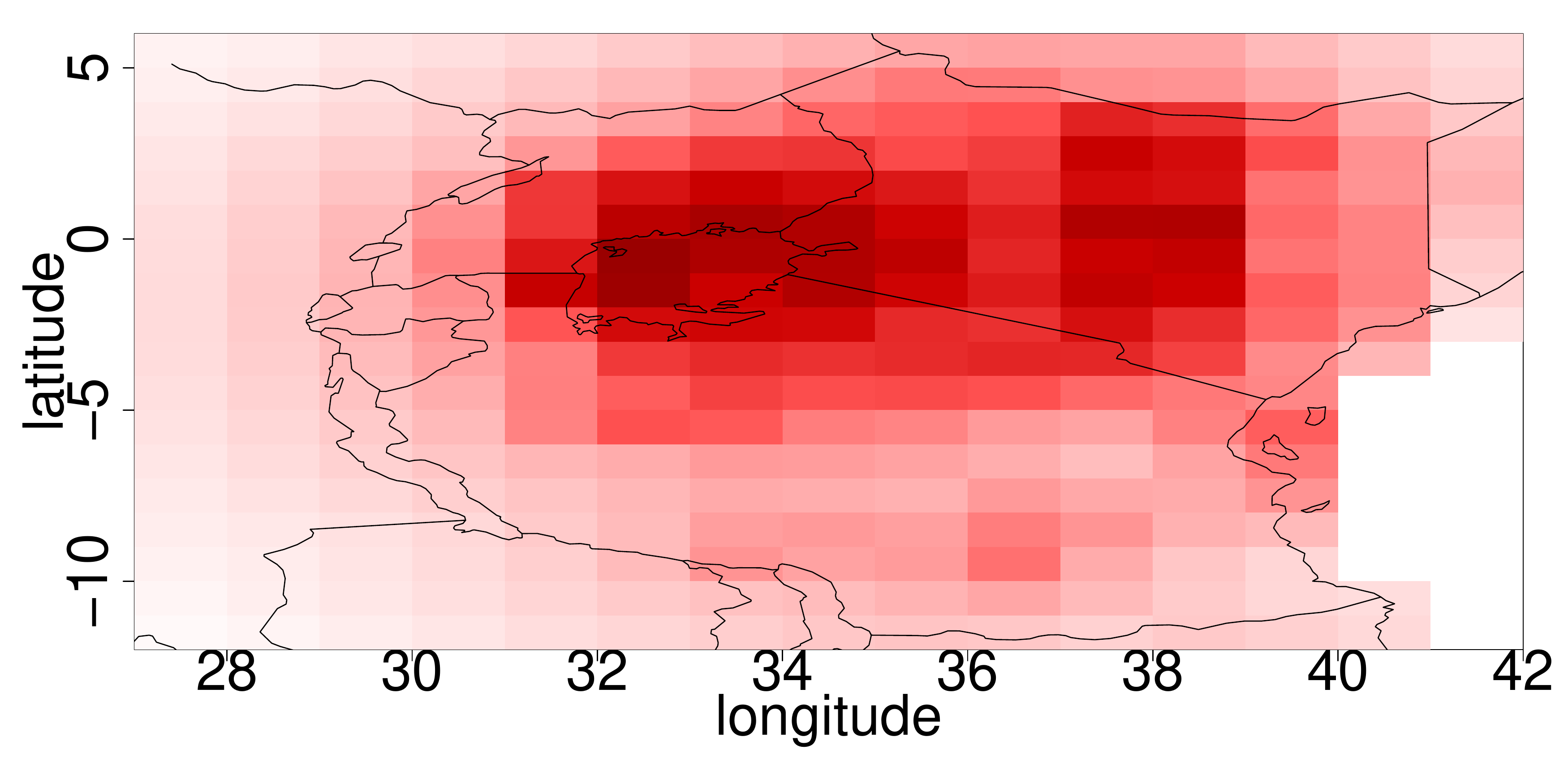}
	\includegraphics[scale=0.2]{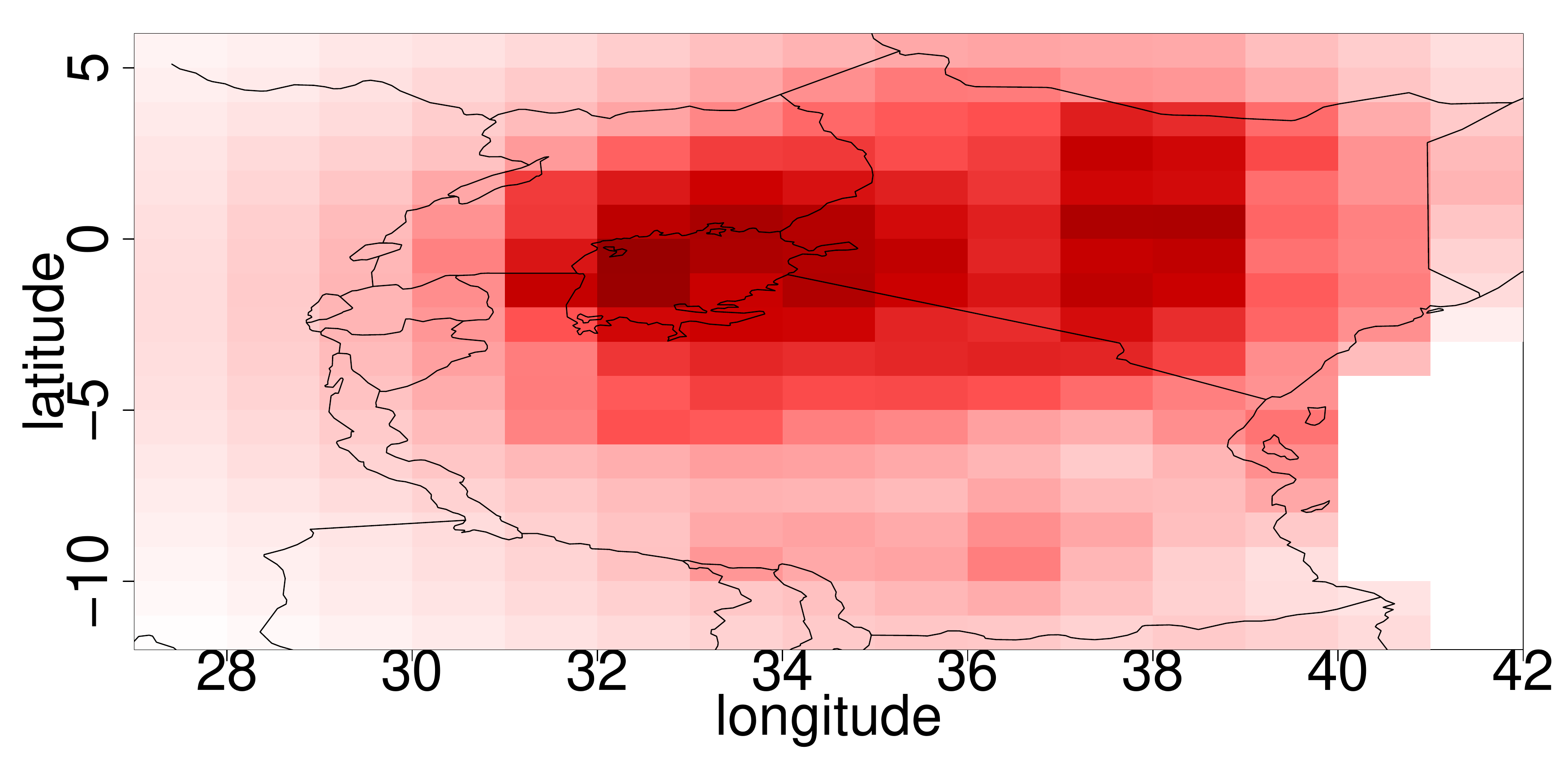}\\
	\includegraphics[scale=0.2]{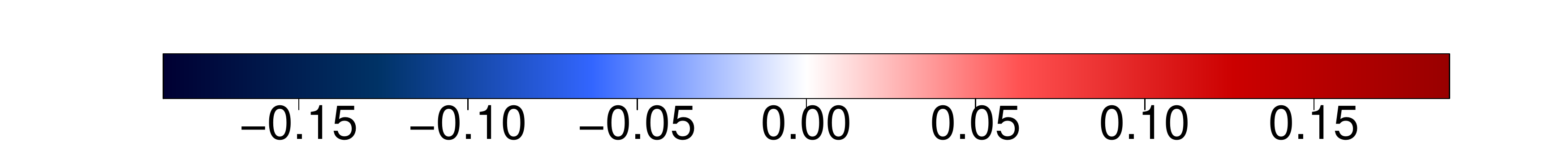}\\
	\caption{{Estimated first coupled} patterns of SST ($\hat{u}_1(\cdot)$) and precipitation ($\hat{v}_1(\cdot)$) from MCA and SpatMCA, where the gray regions correspond to the land with no SST data.} 
	\label{fig:mca_real}
\end{figure}

\begin{figure}\centering
	
	\begin{tabular}{cc}
		MCA& SpatMCA\\
		\includegraphics[height=6cm, width=7 cm]{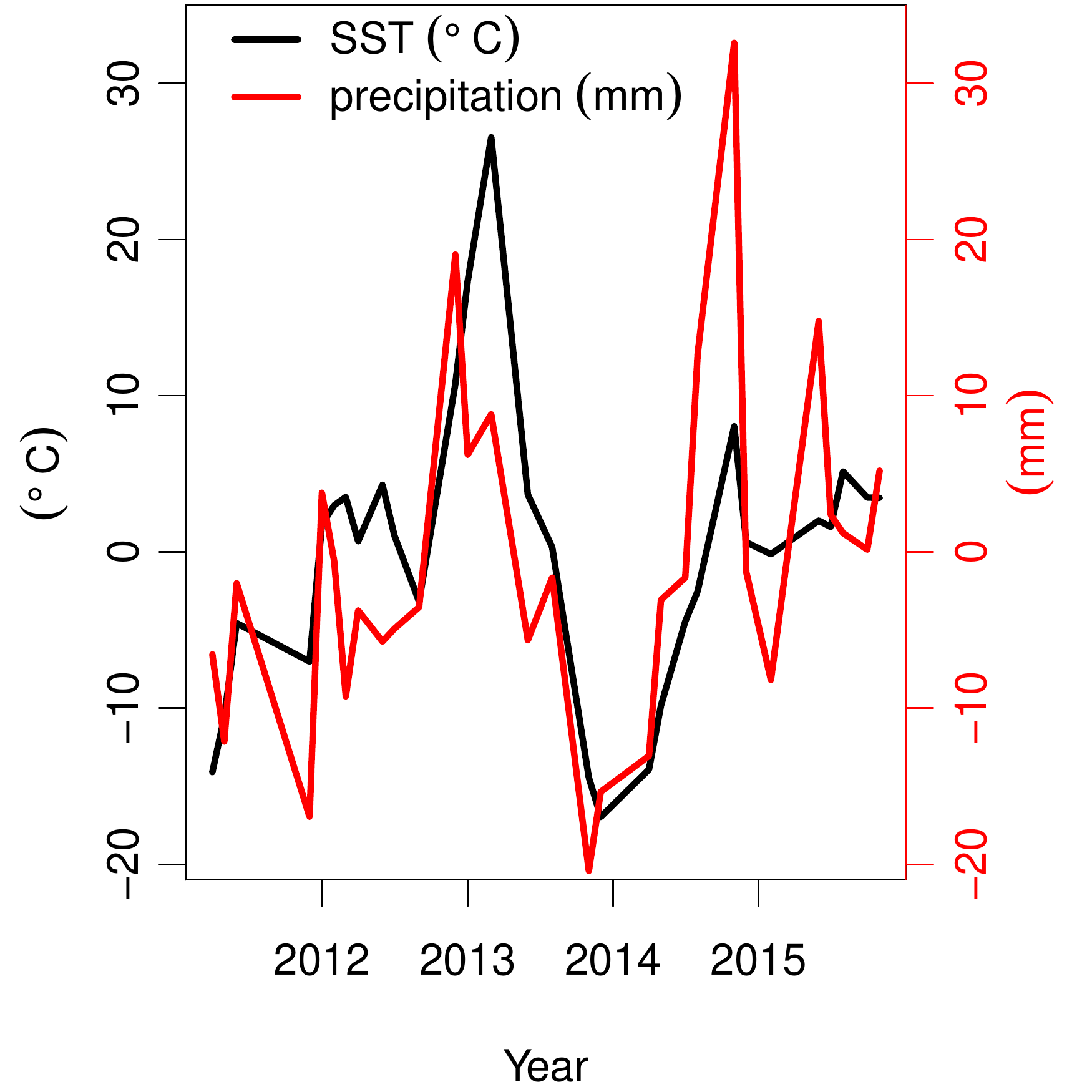}&
		\includegraphics[height=6cm, width=7 cm]{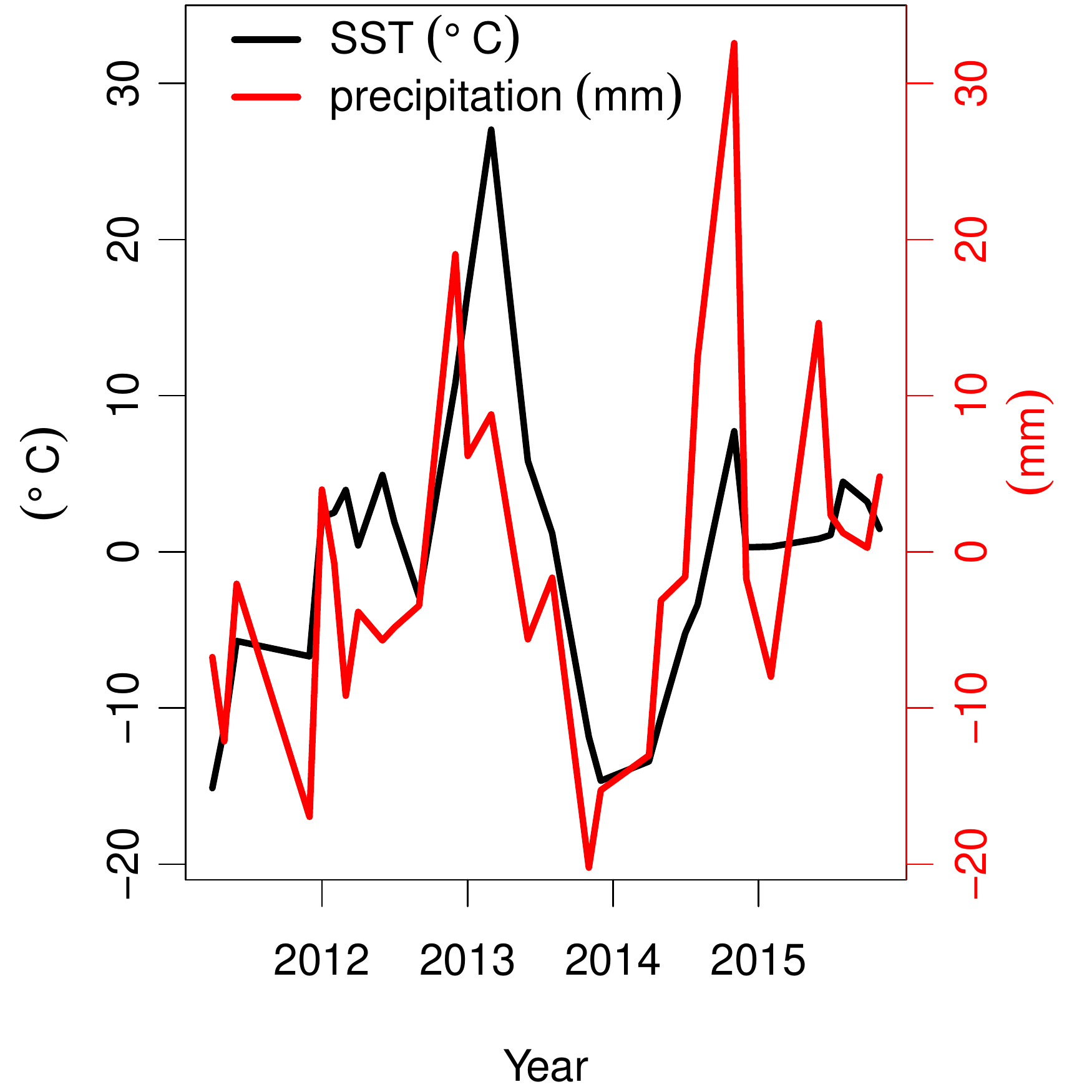}
	\end{tabular}
	\caption{{Time} series of the first  maximum covariance {variables of} SST and precipitation {obtained from MCA and SpatMCA}.}
	\label{fig:mca_ts}
\end{figure}

The best CV values with respect to $K$ for both methods are shown in Figure~\ref{fig:cv_real}. Clearly, both methods selected $\hat{K} = 1$. Figure~\ref{fig:mca_real} shows the first dominant coupled patterns of  SST and precipitation obtained from SpatMCA and MCA. While both methods produce similar patterns, the SST pattern obtained by MCA is much noisier. Figure~\ref{fig:mca_ts} shows {two} time series of the first maximum covariance variables, $\{\hat{\bm{u}}'_1\bm{Y}_{11},\dots,\hat{\bm{u}}'_1\bm{Y}_{1n}\}$ and $\{\hat{\bm{v}}'_1\bm{Y}_{21},\dots,\hat{\bm{v}}'_1\bm{Y}_{2n}\}$, which are the projections of the training data $(\bm{Y}_{j1},\dots,\bm{Y}_{jn})$ for $j=1,2$, onto $\hat{\bm{u}}_1$ and $\hat{\bm{v}}_1$, respectively. As shown in the figure, the first {maximum covariance variables of} SST and precipitation  are highly correlated. Indeed, the Pearson correlation coefficient between {these} two series is 0.59 for SpatMCA and 0.63 for MCA, showing the {importance} of these patterns. 

\begin{figure}
	\centering
	\includegraphics[height=5cm, width=6 cm]{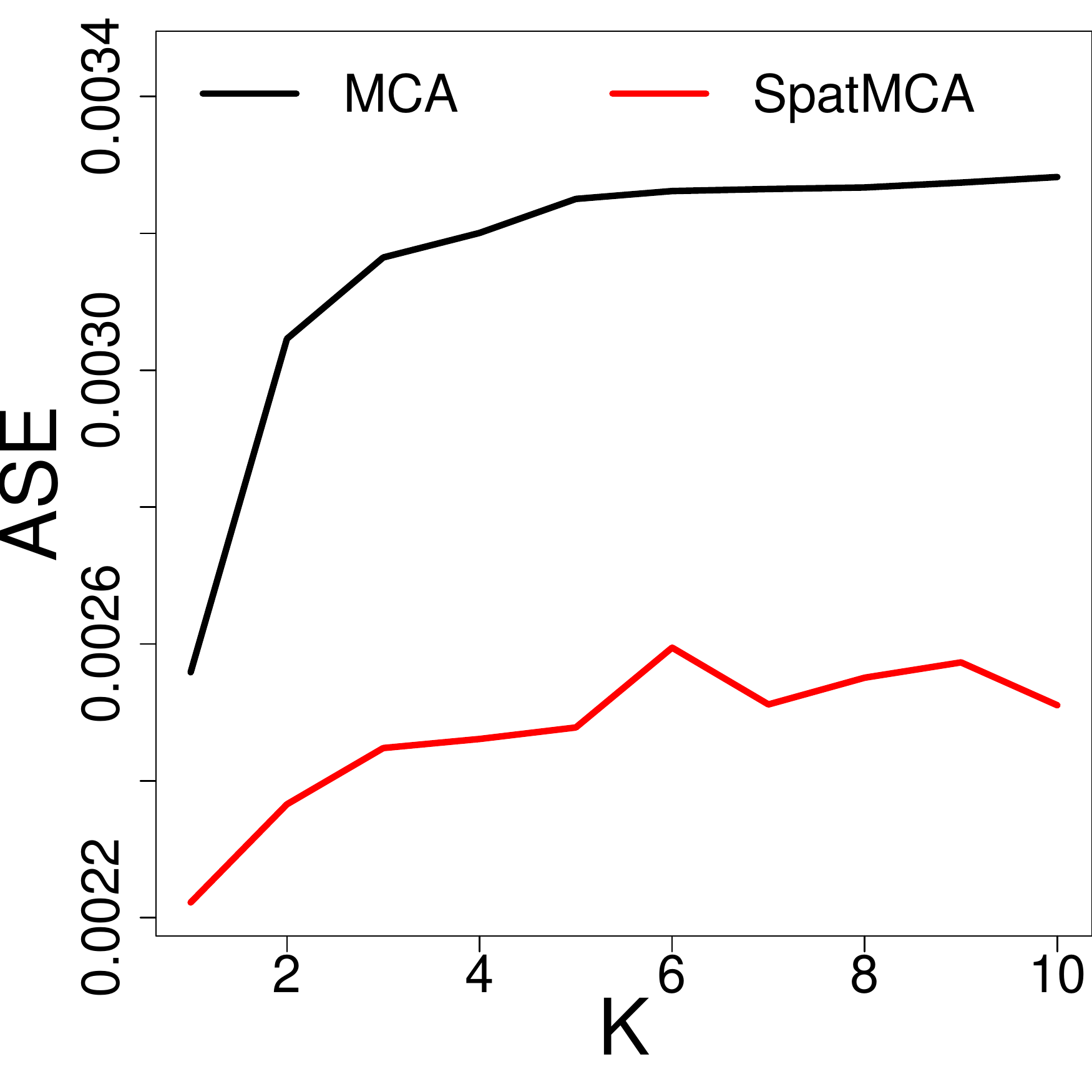}\\
	\caption{Average squared errors of cross-covariance matrix estimates with respect to $K$ for SpatMCA and MCA.}
	\label{fig:mse}
\end{figure}

We further used the validation data to compare the performance between SpatMCA and MCA in terms of the average squared error (ASE),
${\mathrm{ASE}} = \frac{1}{p_1p_2}\|\bm{S}^{v}_{12} - \hat{\bm{\Sigma}}_{12}\|^2_F$, where $\bm{S}^{v}_{12}$ is the sample cross-covariance matrix of the validation data, 
and $\hat{\bm{\Sigma}}_{12}$ is a generic estimate of $\bm{\Sigma}_{12}$. The resulting ASE for MCA is $2.59\times 10^{-3}$, which is larger than $2.25\times 10^{-3}$ for SpatMCA.
Figure~\ref{fig:mse} shows the ASEs with respect to $K$ for both SpatMCA and MCA, which further demonstrate the superiority of SpatMCA over MCA.


\section*{Acknowledgements}
This research was supported in part by ROC Ministry of Science and Technology grant MOST 103-2118-M-001-007-MY3.

\section*{References}
\bibliographystyle{plainnat}
\bibliography{paper}
\end{document}